\documentclass[sigconf]{acmart}

\usepackage{graphicx} 
\usepackage{hyperref}       
\usepackage{url}            
\usepackage[ruled,vlined,linesnumbered]{algorithm2e}
\usepackage{multirow}

\AtBeginDocument{%
  }

\copyrightyear{2026}
\acmYear{2026}
\setcopyright{cc}
\setcctype{by}
\acmConference[MM '26]{Proceedings of the 34th ACM International Conference on Multimedia}{November 10--14, 2026}{Rio de Janeiro, Brazil}
\acmBooktitle{Proceedings of the 34th ACM International Conference on Multimedia (MM '26), November 10--14, 2026, Rio de Janeiro, Brazil}
\acmDOI{10.1145/3767308.3835147}
\acmISBN{979-8-4007-2213-4/2026/11}
\begin{document}

\title{MEC-Patch: Visible-Infrared Cross-Modal Adversarial Attack Driven by Intrinsic Material Emissivity Laws}

\author{Zhixiang Huang}
\affiliation{
\institution{Northwestern Polytechnical University}
\city{Xi'an}
\country{China}
}
\email{huangzhix@mail.nwpu.edu.cn}

\author{Xinbo Nie}
\affiliation{
\institution{Northwestern Polytechnical University}
\city{Xi'an}
\country{China}
}
\email{nxb0813@mail.nwpu.edu.cn}

\author{Wenxuan Wang}
\authornote{Corresponding authors}
\affiliation{
\institution{Northwestern Polytechnical University}
\city{Xi'an}
\country{China}
}
\affiliation{
\institution{Shenzhen Research Institute of Northwestern Polytechnical University }
\city{Shenzhen}
\country{China}
}
\email{wxwang@nwpu.edu.cn}

\author{Lu Yang}
\affiliation{
\institution{Raytron Technology Co.,Ltd}
\city{Xi'an}
\country{China}
}
\email{lu.yang@mail.nwpu.edu.cn}

\author{Xin Li}
\affiliation{
\institution{City University of Hong Kong}
\city{Hong Kong}
\country{China}
}
\email{lixink@mail.nwpu.edu.cn}

\author{Xuelin Qian}
\authornote{Corresponding authors}
\affiliation{
\institution{Northwestern Polytechnical University}
\city{Xi'an}
\country{China}
}
\affiliation{
\institution{Shenzhen Research Institute of Northwestern Polytechnical University }
\city{Shenzhen}
\country{China}
}
\email{xlqian@nwpu.edu.cn}

\author{Peng Wang}
\affiliation{
\institution{Northwestern Polytechnical University}
\city{Xi'an}
\country{China}
}
\email{peng.wang@nwpu.edu.cn}

\renewcommand{\shortauthors}{Huang et al.}

\renewcommand{\footnotetext}[2][]{}

\begin{abstract}
With the widespread deployment of visible-infrared multimodal perception systems in safety-critical domains such as autonomous driving, evaluating their cross-modal adversarial robustness has become increasingly vital. However, existing approaches exhibit significant limitations in approximating the intrinsic laws of imaging. Most studies either focus on a single modality, failing to bypass cross-modal verification, or simplify infrared modeling into heuristic pixel-intensity distributions, neglecting the impact of ambient temperature fluctuations on adversarial stability. To bridge this gap, this paper proposes MEC-Patch, a cross-modal adversarial attack framework driven by intrinsic physical laws. By leveraging the Stefan–Boltzmann Law, we establish a physics-grounded cross-spectral mapping that explicitly links material emissivity to thermal radiation. Building on this formulation, we reveal that, under a fixed emissivity distribution, ambient temperature variations induce consistent global scaling while preserving relative emissivity-induced contrast. We exploit this property to construct temperature-robust adversarial perturbations whose discriminative patterns remain stable in the infrared modality, thereby fundamentally mitigating environmental sensitivity. Furthermore, we employ the physics-constrained NSGA-II algorithm to synergistically optimize the material-distribution-based patch parameters effective across both modalities, while enhancing generalization through a Dynamic Adversarial Resampling (DAR) strategy. Experimental results demonstrate that MEC-Patch effectively deceives state-of-the-art multimodal detectors and exhibits high robustness within high-fidelity, physically-consistent, and multi-scene simulation environments. This research provides a physical-law-driven perspective for the security assessment of multimodal perception systems.
\end{abstract}



\maketitle

\section{Introduction}

In safety-critical visual tasks such as UAV perception, autonomous driving, and all-weather surveillance, multi-modal perception systems integrating Visible (RGB) and Long-Wave Infrared (LWIR) sensors have emerged as the predominant architecture for ensuring operational reliability in complex environments. Fundamentally, LWIR imaging captures the thermal radiation emitted by objects as governed by the Stefan-Boltzmann Law~\cite{vollmer2018infrared}. Its primary advantage lies in the ability to provide stable thermal signal perception by integrating both surface temperature and the intrinsic material property of emissivity, effectively compensating for visible sensor failures under extreme lighting or adverse weather conditions. However, while enhancing environmental robustness, this multi-modal framework simultaneously catalyzes more complex adversarial vulnerabilities. Multi-modal perception algorithms rely heavily on the physical consistency of cross-modal signatures when extracting complementary features~\cite{wei2024physical}. When both visible and infrared modalities are simultaneously subjected to meticulously crafted adversarial perturbations targeting physical attributes, the collaborative perception mechanism is compromised, leading directly to severe performance degradation or misclassification in critical tasks such as object detection. Consequently, conducting in-depth research into cross-modal attacks to identify latent vulnerabilities from a physical perspective is of paramount importance for developing the next generation of highly reliable and secure autonomous perception systems.

While security assessments of multi-modal perception have garnered significant attention, existing adversarial attack methodologies exhibit pronounced limitations when approximating complex physical laws. Early works predominantly focused on single-modal attacks~\cite{liu2020dpatch,xu2020adversarial,jia2025vulnerability}, which struggle to bypass the cross-modal consistency verification of fusion systems. Current cross-modal research primarily adopts a "temperature-control" perspective to construct infrared signatures, involving either active modulation via heating or cooling, or passive modulation by leveraging differential spectral absorption to generate temperature gradients~\cite{zhu2022infrared,wei2023hotcold,long2025cdupatch}. However, this approach reveals fundamental flaws in complex environments: since temperature ($T$) is an extrinsic state variable highly susceptible to external interference, its infrared manifestation is heavily contingent upon ambient lighting, wind speed, atmospheric fluctuations, and the state of the target itself. Consequently, the multi-modal physical consistency of such attacks collapses across diverse scenarios, leading to poor generalization under environmental changes~\cite{liu2024paa}. Existing methods~\cite{wei2023unified,hu2025touap,kim2022map} generally overlook the intrinsic material property of emissivity ($\epsilon$). According to the Stefan–Boltzmann Law, under a fixed emissivity distribution, ambient temperature variations induce consistent global scaling of thermal radiation while preserving relative emissivity contrast. This property provides a principled foundation for moving beyond temperature-dependent modeling toward emissivity-driven formulations, enabling structurally stable cross-modal perturbations under environmental fluctuations.

\begin{figure}[t]
  \centering
  \hspace*{-0.5cm}
  \includegraphics[width=0.5\textwidth]{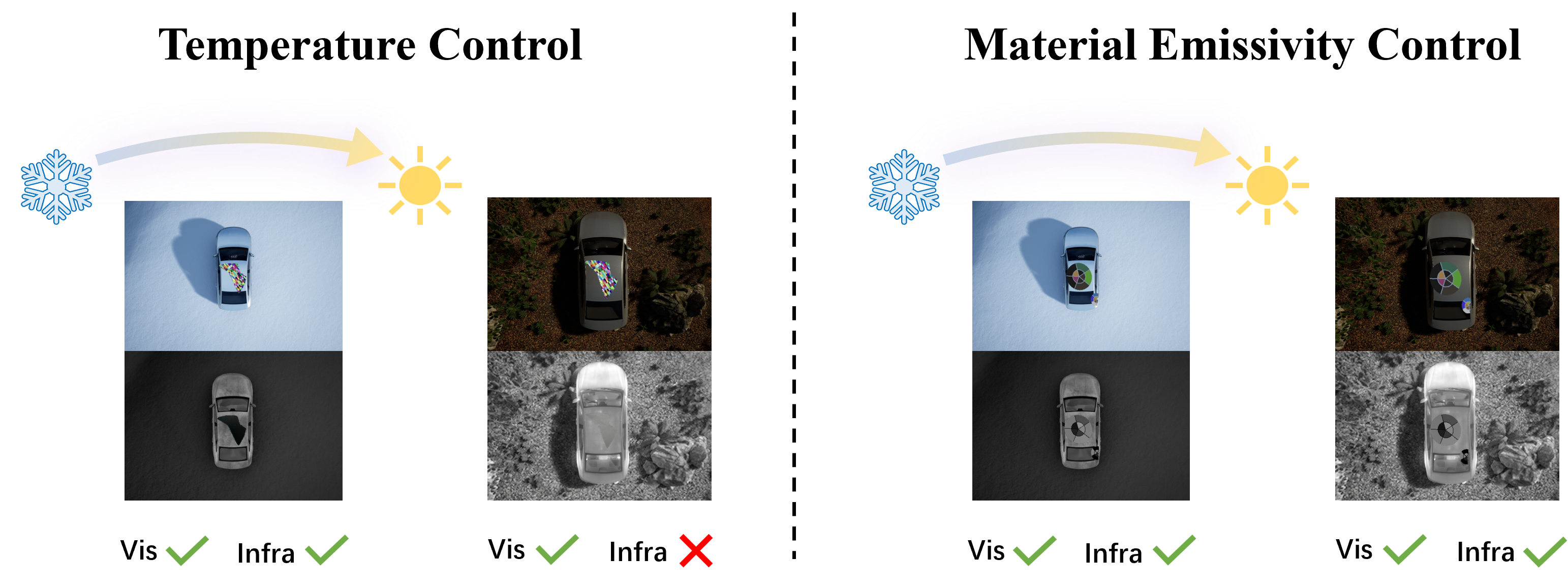} 
  \vspace{-0.1in}
  \caption{Comparison of cross-scene attack robustness. Temperature control fails in infrared (\textcolor{red}{$\times$}) under thermal variations, whereas our Material Emissivity Control (MEC) maintains stable effectiveness (\textcolor{green}{$\checkmark$}) across diverse environments by regulating intrinsic material properties.}
  \label{fig:intro}
  \vspace{-0.1in}
\end{figure}

To this end, this paper introduces MEC-Patch, a Material Emissivity Controlled visible–infrared adversarial attack framework. Instead of manipulating unstable thermal states, MEC-Patch shifts the adversarial space to intrinsic material properties by constructing patches with diverse emissivity and visible color characteristics. Under this formulation, ambient temperature variations primarily affect absolute radiation intensity, while preserving the spatial structure of perturbations, thereby enhancing cross-modal consistency. Despite this advantage, three key challenges remain: (1) effectively establishing the physical mapping between material-specific emissivity and infrared radiation, while simultaneously exploiting surface color to achieve coordinated multi-modal attacks, is a primary hurdle for ensuring inter-modal consistency; (2) emissivity-based patch design involves discrete material selection, spatial configuration, and geometry, leading to a high-dimensional, non-continuous search space that limits gradient-based optimization~\cite{wei2024unified}; (3) although emissivity provides a stable perturbation distribution, environmental fluctuations across multiple scenarios can still impact the overall generalization, necessitating further research to maintain robust cross-modal consistency under diverse physical constraints.

To address these challenges, we propose the MEC-Patch framework grounded in physical material properties, leveraging the intrinsic differences in emissivity and visible-light coloration across various industrial materials (as illustrated in Figure~\ref{fig:intro}) to collaboratively optimize multi-modal attack patches. Through quantitative analysis of the thermodynamic characteristics of diverse materials, we construct patterns with specific infrared radiation signatures via spatially distributed material variations, ensuring inherent cross-modal consistency in physical scenarios. Building upon this foundation,we introduce a discrete spatial topology based on a modular 'Dartboard' partitioning scheme, which transforms the patch geometry into a discretized polar-coordinate representation. By encoding candidate materials as discrete IDs within this topological constraint, we develop a physics-constrained NSGA-II algorithm to map these discrete search spaces into optimizable chromosomal genes, enabling efficient collaborative evolution of dual-modal attack patterns. Furthermore, to ensure the robustness of attack patches under varying environments, we propose a Dynamic Adversarial Resampling Strategy (DAR) strategy, which significantly enhances the generalization capability of attack patterns across complex multi-scenario settings through dynamic sampling and adversarial resampling of environmental temperature, illumination, and observation poses. Our main contributions are summarized as:
\begin{itemize}
\item We propose MEC-Patch, a physics-grounded attack framework that maps emissivity to thermal radiation via the Stefan–Boltzmann Law and incorporates a discrete geometric parameters into a physics-constrained optimization, preserving perturbation structure and enhancing visible-infrared cross-modal physical consistency.
\item We introduce the Dynamic Adversarial Resampling strategy based on multi-scenario sampling optimization to address the weak generalization of targets across environments, where high-fidelity multi-scenario data verify the robustness of adversarial patches under various environmental conditions.
\item We validate MEC-Patch on multiple datasets, demonstrating physically consistent attacks with superior attack success rates against state-of-the-art multimodal detectors, and establishing a new benchmark for physics-driven security evaluation.
\end{itemize}

\section{Related Work}
\subsection{Uni-modal Adversarial Attacks}

Early research on adversarial attacks primarily focused on the visible spectrum. Existing uni-modal attacks can be broadly categorized into appearance-based texture optimization and geometry-driven transformation. During the digital simulation phase, texture-based frameworks such as Naturalistic Patch~\cite{hu2021naturalistic} and ACTIVE~\cite{suryanto2023active} were developed to generate patches with natural textures and universal transferability within digital environments. In contrast, rendering-based approaches such as MeshAdv~\cite{abdelfattahadversarial} and FCA~\cite{wang2022fca} utilized differential rendering techniques to map perturbations onto 3D vehicle models, systematically simulating environmental constraints such as illumination and viewpoints.

As attack methodologies evolved, attention-disruption and camouflage schemes integrating physical stickers and geometric modeling were proposed. DTA~\cite{suryanto2022dta} established a mapping between digital models and physical manifestations through a differential transformation network, demonstrating the feasibility of the digital-to-physical transition. CAC~\cite{duan2021learning} and DAS~\cite{wang2021dual} implemented full-view attacks against vehicle detection by disrupting model attention via adversarial camouflage. For the infrared spectrum, recent studies have begun to explore modality-specific physical features. LMBC~\cite{hu2024physically} designed a segmented transmissive barcode structure, inducing classifier misjudgment by optimizing the sparsity and rotation angles of the stripes. Furthermore, PC-PA~\cite{zhang2024pattern} introduced physical environmental variables into digital optimization through pattern loss loop simulation, significantly enhancing attack effectiveness against high-altitude perception platforms. However, these uni-modal strategies are inherently vulnerable to cross-modal verification in fusion-based systems, as they fail to maintain a synchronized adversarial presence across the electromagnetic spectrum.

\subsection{Visible-infrared Cross-modal Attacks}
To address the requirements of all-weather perception, the research focus is shifting toward visible-infrared cross-modal attacks encompassing both visible and infrared domains. The fundamental discrepancy between the visible spectrum, which relies on surface reflectance and ambient illumination, and the infrared spectrum, governed by intrinsic thermal radiation and material emissivity, poses a significant challenge for establishing cross-modal consistency. At the level of algorithmic synergy and feature mapping, ACAttack~\cite{xiang2025acattack} proposed a spatio-temporal joint attack loss for cross-modal tracking tasks, achieving continuous interference in dynamic scenes by decoupling modal features. CDU-Patch~\cite{long2025cdupatch} attempted to establish a mapping transformation between specific material colors and infrared characteristics through mathematical modeling, utilizing multi-scale cropping strategies to enhance patch robustness under UAV perspectives.Regarding physical implementation and material application, MoXAttack~\cite{han2025evolutionary} proposed a dual-modal synergistic optimization algorithm that alters thermal radiation distribution by adjusting the distribution and thickness of thermal insulation materials, while simultaneously modifying surface textures to interfere with visible light detection. PAIP~\cite{wei2023physically} utilized an aggregation method to optimize the shape and position of low-emissivity insulation materials on target surfaces. Additionally, to improve geometric adaptability, UNIAP~\cite{wei2023unified} introduced a novel boundary-constrained shape optimization method, enhancing the transferability of patches on complex targets through iterative evaluation.

However, most visible-infrared cross-modal attacks optimize in continuous spaces or rely on active temperature control, overlooking the discrete nature of physical materials and intrinsic cross-spectral coupling. In contrast to such temperature-dependent heuristics, MEC-Patch ensures physical plausibility and cross-spectral consistency. Furthermore, the proposed Dynamic Adversarial Resampling (DAR) strategy enhances robustness under discrete physical constraints, enabling stable performance across diverse environments.

\begin{figure*}[htbp]
  \centering
  \includegraphics[width=\textwidth]{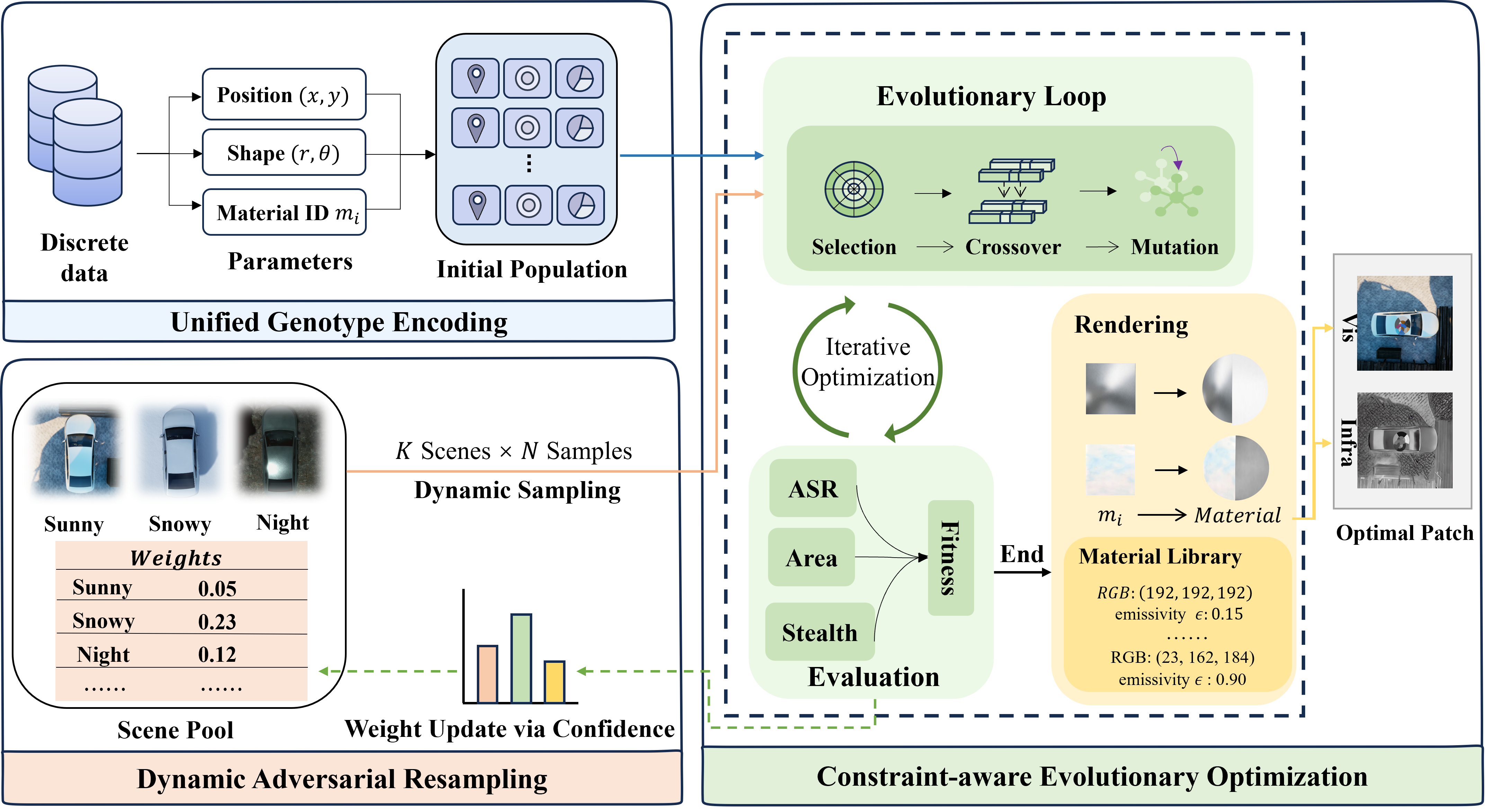}
  \caption{Overview of the MEC-Patch framework. The system consists of three synergetic modules: (1) Unified Genotype Encoding, which maps discrete material and geometric parameters into a chromosomal initial population; (2) Constraint-aware Evolutionary Optimization, where candidate genotypes undergo an iterative loop of selection, crossover, and mutation and are evaluated across three objectives—Attack Success Rate (ASR), Stealth, and Area—until the final parameters are transformed into visualized adversarial patches through a physics-based Rendering process upon convergence; and (3) Dynamic Adversarial Resampling (DAR), which adaptively updates the weights of diverse environments in the Scene Pool based on confidence feedback. Through this bi-level optimization, the framework yields an Optimal Patch that maintains high cross-modal consistency and robustness across both visible and infrared modalities.}
  \Description{}
  \label{fig:pipeline}
  \vspace{-0.05in}
\end{figure*}

\section{Methodology}

\subsection{Preliminary}

\noindent\textbf{Multimodal Object Detection Task.} Consider a pre-trained multimodal object detector $f(\cdot)$ that takes a pair of registered images as input: a visible light image $X_{vis} \in \mathbb{R}^{H \times W \times 3}$ and a long-wave infrared (LWIR) image $X_{ir} \in \mathbb{R}^{H \times W}$. Our objective is to generate a universal adversarial patch $P(\mathcal{G})$ mapped from a genotype encoding $\mathcal{G}$. The attack process aims to minimize the confidence score of the detector regarding the target class. Over a distribution of multiple scenes $\mathcal{S}$, the optimization problem is formulated as:
\begin{equation}
\min_{\mathcal{G}} \mathbb{E}_{s \in \mathcal{S}} [f(X_{vis} \oplus P(\mathcal{G}), X_{ir} \oplus P(\mathcal{G}))]
\label{eq:optimization}
\end{equation}
where $\oplus$ denotes the spatial projection and blending operation, and $\mathcal{G}$ encompasses the geometric, topological, and material parameters of the patch.

\noindent\textbf{Physical Fundamentals of Thermal Radiation.} The core of achieving effective infrared attacks lies in simulating the thermal emission characteristics of real-world materials. According to the Stefan-Boltzmann Law, the total radiant power per unit area $j$ emitted by the surface of an object is defined as:
\begin{equation}
j = \epsilon \cdot \sigma \cdot T^4
\label{eq:stefan_boltzmann}
\end{equation}
where $\sigma \approx 5.67 \times 10^{-8} \, W/(m^2 \cdot K^4)$ is the Stefan-Boltzmann constant and $T$ is the absolute thermodynamic temperature. In most outdoor driving scenarios, the ambient temperature $T$ can be assumed to be relatively stable within a short duration. Consequently, the variation in pixel intensity (\emph{i.e.}, contrast) captured by infrared sensors is primarily dictated by the emissivity ($\epsilon$) of the material surface. While full radiative transfer includes environmental and solar reflections, under typical driving conditions ($  T_{\mathrm{obj}} > T_{\mathrm{env}}  $) self-emission governed by the Stefan–Boltzmann law remains dominant. The inverse relationship between self-emission and reflection naturally preserves relative emissivity contrast, consistent with infrared stealth engineering practices. Based on this physical rationale, we map the material indices within the genotype to specific emissivity values, thereby ensuring the physically-consistent visual appearance of the adversarial patch in the infrared spectrum.

\noindent\textbf{Framework Overview.}
To operationalize this physical mapping, MEC-Patch (Fig.~\ref{fig:pipeline}) encodes discrete material and geometric parameters into a unified genotype. geometric parameters into a unified genotype encoding. Under these physical constraints, a specialized evolutionary algorithm is employed to search the non-convex space, while a Dynamic Adversarial Resampling (DAR) strategy is incorporated to enhance the robustness of the optimized patches across diverse environments.

\subsection{Unified Genotype Encoding}

\noindent\textbf{Parameterized Representation and Structural Constraints.} 
To establish a deterministic mapping between the continuous heuristic search space and discrete physical manufacturing requirements, this study proposes a unified genotype encoding scheme that represents the adversarial patch as a multidimensional chromosomal vector $\mathbf{g} = [\mathbf{g}_{pos}, \mathbf{g}_{sha}, \mathbf{g}_{mat}]$. Specifically, the position genotype $\mathbf{g}_{pos} = [x_{rel}, y_{rel}]$ utilizes normalized relative coordinates to define the center of the patch within the target vehicle's bounding box. This representation achieves the decoupling of the patch's spatial coordinates from absolute image pixels, ensuring that the adversarial perturbation maintains semantic localization constancy during target scaling or displacement. Regarding the morphometric design, the shape genotype $\mathbf{g}_{sha}$ adopts a parameterized polar coordinate system to balance attack flexibility with physical constraints. Its macroscopic contour is defined by the polar equation:
\begin{equation}
r(\theta) = \sqrt{(a \cos\theta)^2 + (b \sin\theta)^2}
\label{eq:polar_contour}
\end{equation}
where the semi-axis parameters $a$ and $b$ enable a continuous evolution from standard circles to proportional ellipses. Considering that the complex and irregular edge geometries in traditional adversarial examples often lead to uncontrollable errors during field cutting and splicing, we introduce a modular "Dartboard" topological structure. This structure discretizes the interior of the closed shape via angular sectors $N_\theta$ and radial rings $N_r$. This design reduces reliance on high-precision cutting while enabling multi-material composition within a single perturbation, enhancing spatial texture complexity and ensuring fabrication accuracy. These encoded genotypes form the initial population for the subsequent evolutionary optimization.

\subsection{Constraint-aware Multi-objective Evolutionary Optimization}
\label{sub_opt}

\noindent\textbf{Evolutionary Logic via NSGA-II.}
Upon establishing the unified genotype, the pivotal challenge shifts toward seeking optimal solutions within a non-convex search space spanned by position, shape, and discrete material indices. Given that the index-based nature of the material genotype $\mathbf{g}_{mat}$ renders the rendering pipeline highly non-linear and non-differentiable, traditional gradient-based optimization strategies frequently fail to converge. To address this, we incorporate the Non-dominated Sorting Genetic Algorithm II (NSGA-II) as a heuristic search engine. In each evolutionary iteration $t$, the algorithm maintains a population of candidate genotypes $\mathcal{P}_t = \{\mathbf{g}_1, \dots, \mathbf{g}_N\}$. Through swarm intelligence-based operators, including tournament selection, simulated binary crossover, and polynomial mutation, the algorithm explores the Pareto front of material compositions globally, thereby avoiding entrapment in local optima. To evaluate the fitness of these evolved candidates, each genotype must first be transformed into a physically consistent digital representation via a dedicated rendering pipeline.

\noindent\textbf{Cross-modal Rendering and Multi-objective Evaluation.} The robustness of multimodal adversarial attacks fundamentally depends on the physical legitimacy of perturbations across different spectra. Specifically, infrared signatures must originate from the intrinsic thermal properties of materials rather than heuristic intensity distributions. In each iteration, the dual-modal rendering engine acts as a forward operator $\mathcal{R}(\cdot)$ that maps genotypes into the physical domain. Within this engine, the material genotype $\mathbf{g}_{mat} = \{m_1, m_2, \dots, m_n\}$ drives two parallel branches: the visible branch retrieves the material's RGB reflectivity for texture synthesis, while the infrared branch extracts the intrinsic emissivity $\epsilon_i$. Under  thermodynamic equilibrium, the infrared pixel intensity $\mathcal{I}$ is derived by:
\begin{equation}
\mathcal{I} = \mathcal{M}(\epsilon_i \cdot \sigma T^4)
\label{eq:rendering_mapping}
\end{equation}
where $\sigma$ is the Stefan-Boltzmann constant and $\mathcal{M}(\cdot)$ represents the mapping function to the sensor’s digital grayscale range. This methodology supports a passive attack mechanism that operates without active heat sources by exploiting the emissivity contrast between industrial materials. For instance, our library incorporates polished Aluminum ($\text{Al}$), which can be coated to appear dark in the visible spectrum but maintains a characteristic low emissivity ($\epsilon \approx 0.15$), resulting in low-intensity "cold" gray blocks in infrared. Conversely, industrial ceramics are utilized for their high emissivity ($\epsilon \approx 0.90$), yielding high-intensity "hot" signatures regardless of their visible hue. By spatially interweaving these materials within the "Dartboard" topology, MEC-Patch synthesizes cross-spectral texture contradictions—such as patches appearing as a uniform pattern in RGB while exhibiting high-frequency, high-contrast grids in the infrared spectrum. This decoupling of visible textures from thermal radiation patterns effectively disrupts the feature alignment and spatial correlation of multimodal detectors.
Finally, the parameterized patch is projected onto the target via a Thin Plate Spline (TPS) operator and augmented with Alpha blending to simulate surface reflection. This pipeline synthesizes adversarial images that strictly adhere to physical laws and maintain robust cross-modal consistency, bypassing the need for heuristic or environment-sensitive perturbations. Our formulation assumes local thermal equilibrium. Under this condition, material emissivity exhibits only weak temperature dependence, as governed by the Fresnel equations and Kirchhoff’s law, thereby ensuring stable contrast across the range of $-20^\circ$C to $50^\circ$C.

Following the rendering process, a multi-objective fitness function $\Phi(\mathbf{g})$ is formulated to evaluate candidates across three dimensions: Attack Success Rate (ASR), Stealth, and Area:
\begin{equation}
\Phi(\mathbf{g}) = [f_{atk}(\mathbf{g}), f_{stealth}(\mathbf{g}), f_{area}(\mathbf{g})]^T
\label{eq:fitness}
\end{equation}
Specifically, the attacking objective $f_{atk}$ is defined as the joint weighted prediction loss across both visible and infrared branches:
\begin{equation}
\begin{split}
f_{atk}(\mathbf{g}) = \mathbb{E}_{s \in \mathcal{S}} \Big[ \lambda_{vis} & \mathcal{L}(f(\mathcal{R}_{vis}(\mathbf{g}, s)), y) \\
& + \lambda_{ir} \mathcal{L}(f(\mathcal{R}_{ir}(\mathbf{g}, s)), y) \Big]
\end{split}
\label{eq:attack_objective}
\end{equation}
where $\mathcal{L}$ denotes the prediction loss, $\mathcal{R}_{vis}$ and $\mathcal{R}_{ir}$ represent the visible and infrared rendering operators, respectively, and $s$ denotes the scene state sampled by the DAR strategy(see Section ~\ref{sec:dar}. The stealthiness objective $f{stealth}$ minimizes the color distribution distance to the background, while $f_{area}$ penalizes excessive patch size to maintain physical practicality. Through non-dominated sorting and crowding distance, NSGA-II preserves population diversity among these conflicting objectives, yielding a Pareto-optimal set. Upon the completion of iterative optimization, the candidate with the highest fitness is rendered to generate the final Optimal Patch.

\subsection{Dynamic Adversarial Resampling Strategy}
\label{sec:dar}

To ensure the evolutionary process achieves distributional robustness rather than instance-specific success, we propose the Dynamic Adversarial Resampling (DAR) strategy. We conceptualize this framework as a bi-level adversarial game: while the inner loop utilizes NSGA-II to explore optimal material genotypes, the outer loop (DAR) dynamically reconstructs the fitness landscape. As illustrated in the Dynamic Adversarial Resampling block, DAR acts as a non-stationary wrapper that directly intervenes in the evaluation phase of each evolutionary generation.

\noindent\textbf{Intra-scene Spatial Invariance.} To simulate local sensing diversity, the algorithm performs parallel evaluation by sampling $N$ different spatial configurations within a specific environment. By introducing stochastic background interference, this mechanism transforms the deterministic fitness value into a stochastic expectation, compelling the inner-loop engine to learn universal discriminative features on the vehicle surface rather than over-fitting to specific background pixels.

\noindent\textbf{Inter-scene Adversarial Learning.} To address population degradation caused by imbalanced attack difficulty across heterogeneous environments, we maintain a Scene Weight Pool containing diverse environments such as Sunny, Snowy, and Night. In each generation, the algorithm adaptively selects $K$ representative scenes from the pool based on their current weights $\mathbf{W} = [w_1, \dots, w_S]$ to form a comprehensive evaluation set ($K < S$). Specifically, the attack error $e_s^t$ for scene $s$ is derived from the Confidence Calculation in the Evaluation module:
\begin{equation}
e_s^t = 1 - \frac{1}{M}\sum_{j=1}^M \text{conf}(x_j \oplus P)
\label{eq:attack_error}
\end{equation}
where $\text{conf}(\cdot)$ denotes the confidence score of the target class. A lower error $e_s^t$ indicates a harder scenario (higher residual confidence). Eq.~\eqref{eq:weight_update} therefore assigns larger weights to scenes with lower $e_s^t$. The adaptive update rule for weights is formulated as:
\begin{equation}
w_s^{t+1} = \alpha w_s^t + (1-\alpha) \cdot \frac{\exp(1-e_s^t)}{\sum_{i=1}^S \exp(1-e_i^t)}
\label{eq:weight_update}
\end{equation}
where $\alpha$ denotes the momentum smoothing factor. Under this bi-level framework, higher weights are assigned to challenging scenarios, forcing the optimization process to prioritize high-error distributions. This ensures that the MEC-Patch converges toward a distribution that is robust against varied environmental conditions.

\section{Experiment}

\begin{table*}[htbp]
\centering
\caption{ASR (\%) of different methods against various detectors on DroneVehicle, LLVIP, and ViSDrone datasets.}
\label{tab:asr_main}
\vspace{-0.1in}
\begin{tabular}{lccccccccccccccc}
\hline
\multirow{2}{*}{\textbf{Method}} & \multicolumn{5}{c}{DroneVehicle} & \multicolumn{5}{c}{ViSDrone}& \multicolumn{5}{c}{LLVIP}  \\
\cline{2-16}
& v3 & v5 & v8 & v11 & RCNN & v3 & v5 & v8 & v11 & RCNN & v3 & v5 & v8 & v11 & RCNN \\
\hline
RandomPatch & 11.2 & 8.8 & 2.4 & 1.9 & 8.3 & 13.7 & 10.5 & 5.6 & 4.7 & 11.1 & 16.5 & 13.9 & 9.3 & 7.4  & 14.7\\
TOUAP & 31.6 & 28.2 & 21.0 & -- & 25.7 & 34.8 & 31.0 & 24.2 & -- & 29.3 & 38.6 & 33.7 & 27.5 & --  & 32.8 \\
UNIAP & 46.9 & 41.6 & 33.7 & 30.6 & 38.8  & 49.2 & 44.9 & 37.5 & 32.9 & 42.6  & 52.0 & 47.3 & 40.3 & 38.7 & 45.2  \\
CDUPatch & 70.4 & 66.7 & 57.5 & -- & 63.2 & 71.8 & 68.5 & 60.3 & -- & 65.6 & 73.2 & 73.1 & 63.0 & -- & 68.0  \\
\textbf{MEC-Patch (ours)} & \textbf{72.5} & \textbf{77.7} & \textbf{81.7} & \textbf{73.7} &  \textbf{64.1} & \textbf{72.4} & \textbf{72.0} & \textbf{69.8} & \textbf{69.0} & \textbf{77.2}  & \textbf{80.3} & \textbf{86.1} & \textbf{80.8} & \textbf{72.2} & \textbf{77.1}\\
\hline
\end{tabular}
\vspace{-0.1in}
\end{table*}

\begin{figure*}[htbp]
\centering
\includegraphics[width=\linewidth]{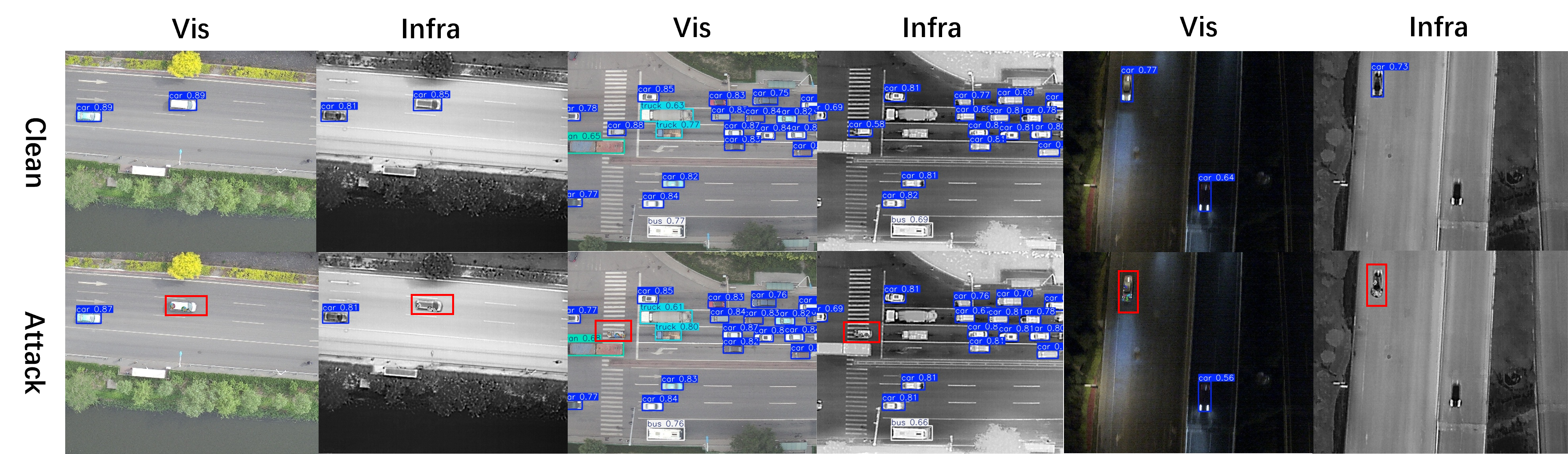}
\vspace{-0.15in}
\caption{Qualitative visualization of detection results on clean images with adversarial patches.}
\label{fig:visualization}
\end{figure*}

\subsection{Experimental Setup}

\noindent\textbf{Datasets.}
Following the experimental protocol in \cite{long2025cdupatch}, we evaluate on three representative visible–infrared (RGB-IR) datasets: DroneVehicle~\cite{sun2022dronevehicle}, LLVIP~\cite{jia2021llvip}, and VisDrone~\cite{zhu2020visdrone}. DroneVehicle contains 28,439 strictly registered UAV-view image pairs under day-to-night conditions. LLVIP focuses on low-light pedestrian detection with precise spatial alignment, while VisDrone provides a large-scale UAV benchmark with complex backgrounds and multi-scale objects. Training sets are used to fine-tune detectors for defense baselines.

\noindent\textbf{Victim Models.}
The evaluation encompasses two mainstream detection paradigms: one-stage detectors, including YOLOv3 (\textbf{v3})~\cite{redmon2018yolov3}, YOLOv5 (\textbf{v5})~\cite{jocher2020yolov5}, YOLOv8 (\textbf{v8})~\cite{jocher2023yolov8}, and the latest YOLOv11 (\textbf{v11})~\cite{yolo11_ultralytics}; and a two-stage detector, Faster R-CNN (\textbf{RCNN})~\cite{ren2015faster}. Following \cite{long2025cdupatch}, all victim models are initialized and fine-tuned to ensure strong detection performance, achieving around 0.9 confidence in benign vehicle detection across scenarios.

\noindent\textbf{Competitors.}
Hot-Cold is tailored for human thermal signatures and shows limited effectiveness on vehicles; MoXAttack lacks sufficient implementation details for faithful reproduction; PAIP is the predecessor of UNIAP. We therefore focus on the most relevant and reproducible baselines under consistent physical constraints. Therefore, we focus our comparisons on the most relevant and reproducible baselines under consistent physical constraints. Four representative methods are selected as competitors: 1) Random Patch, a non-optimized baseline with random initialization of position and texture; 2) TOUAP ~\cite{hu2025touap}, a unified dual-modal attack optimized via pixel-level joint loss; 3) UNIAP ~\cite{wei2023unified}, which generates universal patches using spatial constraints and cross-modal consistency losses; and 4) CDUPatch ~\cite{long2025cdupatch}, a color-driven attack exploiting correlations between RGB color and thermal responses. Experimental settings follow the ~\cite{long2025cdupatch}.

\noindent\textbf{Implementation Details.}
Experiments are conducted in PyTorch, with the inference resolution for all detectors set to $640 \times 640$ pixels. The material library contains $N$ (\emph{e.g.}, 20) representative industrial materials (metals, polymers, coatings), whose infrared emissivity ($\epsilon$) and RGB reflectivity are calibrated from standardized databases to ensure physical realism. During the adversarial evolution phase, the population size for NSGA-II is set to $60$, and the maximum number of generations is fixed at $80$. For the Dynamic Adversarial Resampling (DAR) strategy, the momentum factor $\alpha$ is set to $0.9$, with $K=5$ representative scenes sampled from the pool and $N=3$ spatial configurations sampled for each scene per iteration. The maximum coverage area of the patch is restricted to $30\%$ of the target's bounding box. The primary metric is the Attack Success Rate (ASR), defined as the ratio where the detector fails to predict the correct category with an IoU below the threshold. All experiments are conducted on an NVIDIA GeForce RTX 4090 GPU.

\subsection{Attack Results}

\noindent\textbf{Analysis of Dual-modal Attack Efficacy and Material Property Correlation.}
Table~\ref{tab:asr_main} reports the ASR of our method against various detectors on three benchmark datasets. Results show consistent superiority over all baselines. For example, on DroneVehicle, our method achieves 77.7\% ASR against YOLOv5, significantly outperforming active temperature-control-based attacks. This gain stems from accurate modeling of intrinsic material properties. Unlike conventional thermal simulation methods that rely on heuristic RGB–thermal correlations, our approach introduces material emissivity ($\epsilon$) to model radiation responses consistent with thermodynamic laws under passive equilibrium. Under complex backgrounds, prior methods often suffer from cross-modal feature misalignment, whereas our method preserves perturbation consistency across dual-spectral semantic spaces via emissivity contrasts encoded in the material gene $\mathbf{g}_{mat}$. Overall, these results indicate that modeling intrinsic material properties more effectively disrupts the feature fusion process of dual-modal detectors, providing a principled foundation for physical-world adversarial attacks.

\noindent\textbf{Qualitative Visualization and Structural Effectiveness.}
Figure~\ref{fig:visualization} compares detection results on clean images and targets with adversarial patches. The results show that our patch achieves strong visual stealth while strictly satisfying the $30\%$ area constraint: originally high-confidence vehicles are either missed or suffer severe category shifts. The dartboard-like polar partitioning creates cross-modal texture contradictions by interleaving high- and low-emissivity materials within a limited area, effectively inducing perceptual bias in the detectors. These qualitative observations validate both the partitioned design and the efficiency of the heuristic search in finding globally optimal adversarial configurations under physical constraints. Quantitatively, on DroneVehicle-Far / LLVIP-Near, MEC-Patch attains SSIM = 0.970 / 0.712 and LPIPS = 0.051 / 0.203, outperforming UNIAP (SSIM = 0.952 / 0.615, LPIPS = 0.068 / 0.284), confirming that the emissivity-driven patches maintain high visual naturalness while delivering strong adversarial effectiveness.

\noindent\textbf{Cross-model Transferability Assessment.}
To evaluate the universality of MEC-Patch in black-box settings, we conduct cross-model transferability tests (Table~\ref{tab:transferability}). Results show that patches optimized on a one-stage detector (\emph{e.g.}, YOLOv11) retain a stable 59.9\% ASR when transferred to a two-stage detector (\emph{e.g.}, Faster R-CNN) with a different architecture. This transferability stems from our physics-driven generation process, which is grounded in intrinsic material emissivity laws rather than network-specific gradients, yielding more universal feature response deviations across visible and infrared modalities. These results indicate that physically defined adversarial “genes” generalize beyond specific architectures, posing potential threats to unseen defense systems in real-worlds.

\subsection{Multi-scenario Performance}

\noindent\textbf{Quantitative Evaluation across Multiple Scenarios.}
To verify the cross-scene generalization of the adversarial patches, we conducted a comprehensive comparative study across multiple simulated environments, including snowy fields, deserts, and urban nightscapes. Table~\ref{tab:multi_scenario} reports the ASR of our MEC-Patch and UNIAP methods. The results demonstrate that the proposed method consistently achieves superior performance across multiple scenarios, maintaining a high average ASR of 56.5\% despite the presence of ambient temperatures and complex background.
Unlike existing methods that often suffer from significant performance drops when transitioning between diverse terrains, MEC-Patch exhibits remarkable stability across all tested detectors (ranging from 53.5\% to 59.0\%). This is attributed to our physics-driven modeling approach: by anchoring adversarial features in intrinsic material emissivity ($\epsilon$) rather than transient pixel intensities, and concurrently employing the Dynamic Adversarial Resampling (DAR) strategy to prioritize hard-to-attack environmental distributions, the generated patches can adapt to various environmental distributions without losing their adversarial potency. This cross-scenario consistency underscores the effectiveness of leveraging intrinsic material laws for robust cross-modal attacks.

\begin{table}[t]
\caption{Cross-model transferability results. Adversarial patches optimized on source models tested on target models.}
\label{tab:transferability}
\setlength{\tabcolsep}{3.2mm}{
\begin{tabular}{lccccc}
\hline
\textbf{Detector} & \textbf{v3} & \textbf{v5} & \textbf{v8} & \textbf{v11} & \textbf{RCNN}  \\
\hline
YOLOv3 & \textbf{72.5} & 62.0 & 60.6 & 40.6 & 52.0 \\
YOLOv5 & 59.9 & \textbf{77.6} & 62.7 & 38.9 & 55.6 \\
YOLOv8 & 50.7 & 60.3 & \textbf{81.7} & 36.3 & 49.4 \\
YOLOv11 & 52.9 & 62.2 & 57.1 & \textbf{73.6} & 59.9 \\
Fast-RCNN & 58.8 & 56.7 & 62.7 & 37.8 & \textbf{64.1} \\
\hline
\end{tabular}}
\end{table}

\begin{table}[t]
\centering
\caption{Comparison of ASR (\%) across multiple scenarios.}
\label{tab:multi_scenario}
\setlength{\tabcolsep}{2.8mm}{
\begin{tabular}{lccccc}
\hline
\textbf{Detector} & \textbf{v3} & \textbf{v5} & \textbf{v8} & \textbf{v11} & \textbf{RCNN}  \\
\hline
RandomPatch & 1.2 & 0.8 & 0.4 & 1.0 & 1.4 \\
UNIAP & 21.4 & 18.6 & 15.2 & 17.8 & 19.4 \\
\textbf{MEC-Patch} & \textbf{56.5} & \textbf{59.0} & \textbf{58.5} & \textbf{53.5} & \textbf{55.5} \\
\hline
\end{tabular}}
\end{table}

\begin{figure}[t]
\centering
\includegraphics[width=\linewidth]{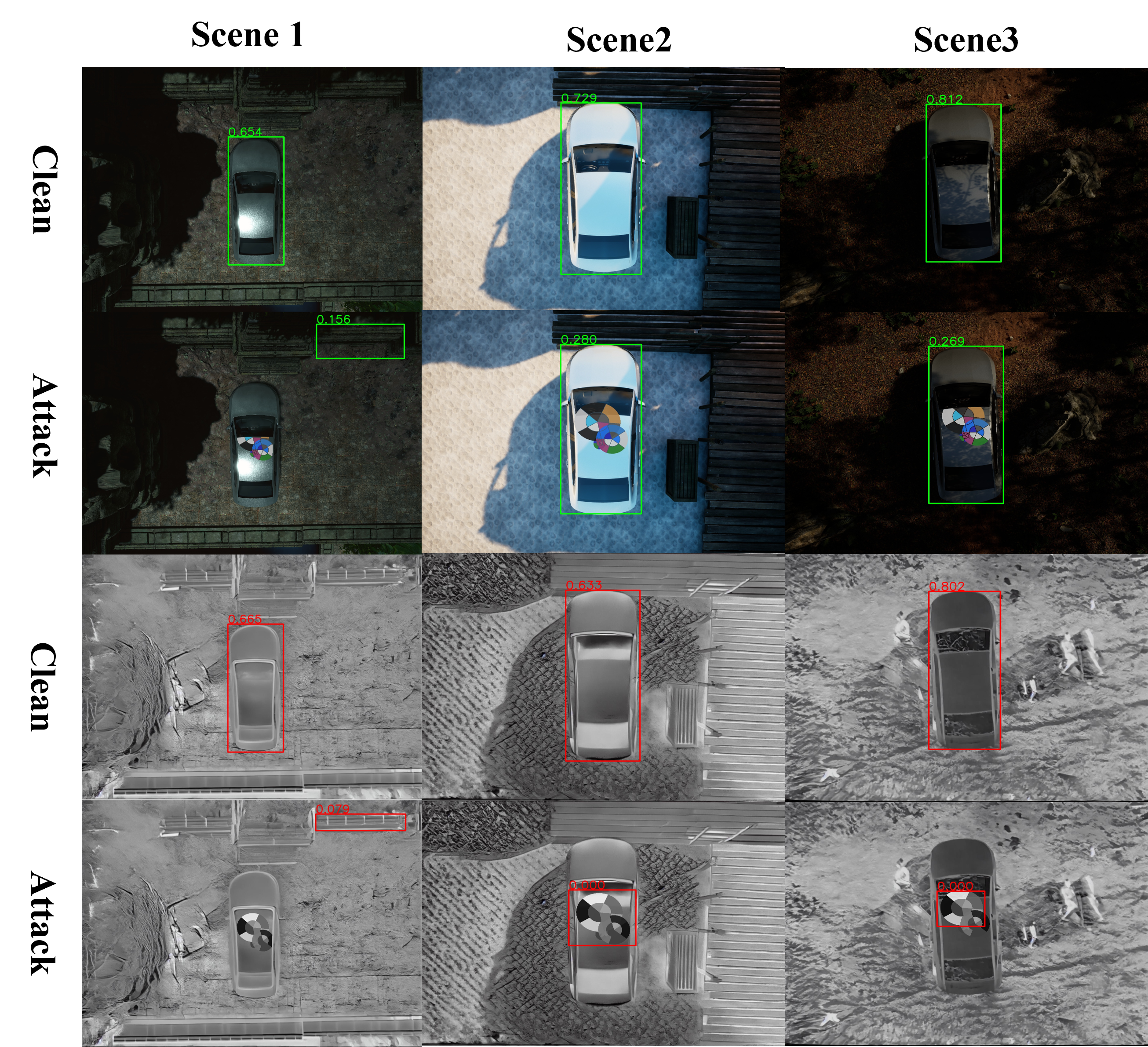}
\caption{Qualitative visualization of detection results across multiple scenarios.}
\label{fig:multi_scene_vis}
\end{figure}

\noindent\textbf{Qualitative Analysis and Modal Consistency across Scenarios.}
Figure~\ref{fig:multi_scene_vis} illustrates the detection results comparison between clean images and targets deployed with adversarial patches across the aforementioned terrains. The visualizations reveal the interference efficacy of the proposed scheme under different environments: in all test scenarios, originally high-confidence targets result in either missed detections or exhibit a drastic decline in confidence scores (drop exceeds 0.4). It is observed that, unlike traditional methods which often suffer from single-modal failure during sudden environmental changes, the patches generated by our method exhibit strong consistency across the dual-spectral multiple scene. This stability ensures that the emissivity-driven adversarial perturbations remain potent across various scene distributions, maintaining high-performance degradation of the target's identifiable features.

\subsection{Ablation Studies}
\label{sec:ablation}

\noindent\textbf{Adversarial Parameters and Structural Constraints.}
Table~\ref{tab:ablation} provides a quantitative breakdown of the contributions of position ($\mathbf{g}_{pos}$), shape ($\mathbf{g}_{shp}$), and material genes ($\mathbf{g}_{mat}$) to the overall attack efficacy. The synergistic optimization of these three components allows MEC-Patch to achieve a peak dual-modal ASR of 64.2\% and 76.8\% on the respective victim models. Specifically:
\textbf{a) Positional Optimization ($\mathbf{g}_{pos}$)}:
This is identified as a critical factor for initial feature disruption. Removing the positional constraint leads to the most significant performance degradation, with infrared ASR dropping to 18.5\% and 15.2\% on YOLOv5 and YOLOv11, respectively. This suggests that adversarial effects are highly sensitive to the spatial semantics and key feature regions of the vehicle surface.
\textbf{b) Morphological and Structural Design ($\mathbf{g}_{shp}$)}:
The "Dartboard" polar partition provides a 20-30\% ASR gain over fixed geometries (\emph{e.g.}, MEC-Patch w/o $\mathbf{g}_{shp}$ using a rectangular template). Its modular design facilitates the practical splicing and assembly of physical materials, reducing manufacturing complexity. Furthermore, compared to stochastic grids, this encoding enables more disruptive pattern compositions that effectively decouple the aligned features across visible and infrared spectra.
\textbf{c) Material Property Modeling ($\mathbf{g}_{mat}$)}:
This serves as the fundamental prerequisite for cross-modal consistency. When material constraints are removed (simulated by the w/o $\mathbf{g}_{mat}$ case using heuristic intensity fitting), the infrared ASR collapses to under 18\%. This contrast proves that while $\mathbf{g}_{pos}$ and $\mathbf{g}_{shp}$ provide geometric freedom, only the Material Emissivity Controlled (MEC) strategy ensures that the generated perturbations remain physical consistency with the Stefan-Boltzmann Law and potent. This adherence to intrinsic material properties ensures that the simulated infrared signatures remain grounded in realistic radiation physics rather than heuristic pixel manipulation.

\begin{table}[t]
\centering
\caption{Ablation study of different components.}
\label{tab:ablation}
\setlength{\tabcolsep}{4mm}{
\begin{tabular}{@{}llc@{}}
\toprule
\multicolumn{1}{c}{Victim} & \multicolumn{1}{c}{Attacker} & ASR (\%) \\ \cmidrule(l){3-3} 
& & Visible / Infrared \\ \midrule
\multirow{5}{*}{YOLOv5} & MEC-Patch w/o $g_{pos}$ & 32.4 / 18.5 \\
& MEC-Patch w/o $g_{shp}$ & 45.8 / 42.1 \\
& MEC-Patch w/o $g_{mat}$ & 36.3 / 17.4 \\
& \textbf{MEC-Patch} & \textbf{64.2 / 76.8} \\ \midrule
\multirow{5}{*}{YOLOv11} & MEC-Patch w/o $g_{pos}$ & 28.7 / 15.2 \\
& MEC-Patch w/o $g_{shp}$ & 41.5 / 38.6 \\
& MEC-Patch w/o $g_{mat}$ & 33.6 / 15.8 \\
& \textbf{MEC-Patch} & \textbf{61.0 / 78.4} \\ \bottomrule
\end{tabular}}
\end{table}

\noindent\textbf{Dynamic Adversarial Resampling Mechanism.}
We evaluate the convergence of MEC-Patch against a uniform sampling baseline. While uniform sampling quickly plateaus due to local optima, the DAR-enhanced population exhibits a distinct late-stage breakthrough. Although DAR maintains lower ASR initially by focusing on hard examples, it ultimately boosts global ASR by approximately 15 percentage points. This confirms DAR as the key driver for overcoming scene imbalance and ensuring robust generalization. Detailed convergence curves are provided in the supplementary material.

\section{Conclusion}
We propose a physics-grounded cross-modal adversarial attack framework for visible–infrared perception security evaluation. Leveraging the Stefan–Boltzmann Law, we map emissivity to infrared radiation to ensure spectral consistency and reduce environmental sensitivity. Coupled with physics-constrained NSGA-II and Dynamic Adversarial Resampling, our method effectively deceives state-of-the-art multimodal detectors. Experiments demonstrate that MEC-Patch achieves superior cross-modal adversarial consistency between visible and infrared spectra. Furthermore, the framework maintains robust attacking potency across varying environmental conditions, revealing latent vulnerabilities in multimodal sensing systems.

\section{Acknowledgments}
This work is supported by the National Natural Science Foundation of China (No. 62502387 \& No.62406252) , the China Postdoctoral Science Foundation (No. BX20250486 \& No. 2025M784419), the Natural Science Basic Research Program of Shaanxi (No. 2025JC-YBQN-861), the Guangdong Basic and Applied Basic Research Foundation (No. 2025A1515011465 \& No. 2025A1515010158) and the Postdoctoral Science Foundation of Shaanxi Province (No. 2025BSHSDZZ105)

\clearpage
\bibliographystyle{ACM-Reference-Format}
\bibliography{main}

@article{wei2024physical,
  title={Physical Adversarial Attack Meets Computer Vision: A Decade Survey},
  author={Wei, Hui and Tang, Hao and Jia, Xuemei and Wang, Zhixiang and Yu, Hanxun and Li, Zhubo and Satoh, Shin'ichi and Van Gool, Luc and Wang, Zheng},
  journal={IEEE Transactions on Pattern Analysis and Machine Intelligence},
  volume={46},
  number={12},
  pages={9797--9817},
  year={2024},
  month={December},
  publisher={IEEE},
  doi={10.1109/TPAMI.2024.3430860},
  issn={1939-3539},
  pmid={39024087}
}

@inproceedings{liu2020dpatch,
  title={Dpatch: An adversarial patch attack on object detectors},
  author={Liu, Xin and Yang, Huan and Liu, Ziwei and Song, Lingxi and Li, Hongyang and Chen, Yiran},
  booktitle={Proceedings of the IEEE Conference on Computer Vision and Pattern Recognition},
  pages={2849--2858},
  year={2020}
}

@inproceedings{xu2020adversarial,
  title={Adversarial T-shirt! Evading person detectors in a physical world},
  author={Xu, Kaidi and Zhang, Gaoyuan and Liu, Sijia and Wang, Quanfu and Lin, Weiyue and Yang, Ming and Chen, Pin-Yu},
  booktitle={European Conference on Computer Vision},
  pages={665--681},
  year={2020},
  publisher={Springer}
}

@article{jia2025vulnerability,
  title={From Vulnerability to Robustness: A Survey of Patch Attacks and Defenses in Computer Vision},
  author={Jia, J and Liu, X and Yang, Y},
  journal={Electronics},
  volume={14},
  number={23},
  pages={4553},
  year={2025},
  publisher={MDPI}
}

@inproceedings{zhu2022infrared,
  title={Infrared invisible clothing: Hiding from infrared detectors at multiple angles in real world},
  author={Zhu, Xiaopei and Hu, Zhuo and Huang, Siyuan and Li, Jianmin and Hua, Xian-Sheng},
  booktitle={Proceedings of the IEEE/CVF Conference on Computer Vision and Pattern Recognition},
  pages={13317--13326},
  year={2022}
}

@inproceedings{wei2023hotcold,
  title={Hotcold block: Fooling thermal infrared detectors with a novel wearable design},
  author={Wei, Hui and Wang, Zibo and Jia, Xiaojun and Zheng, Yinqiang and Liu, Bin},
  booktitle={Proceedings of the AAAI Conference on Artificial Intelligence},
  volume={37},
  number={12},
  pages={15233--15241},
  year={2023}
}

@inproceedings{liu2024paa,
  title={Paa-Tee: A Practical Adversarial Attack on Thermal Infrared Detectors with Temperature and Pose Adaptability},
  author={Liu, X and Wang, Y and Zhang, L and others},
  booktitle={2024 IEEE 23rd International Conference on Trust, Security and Privacy in Computing and Communications (TrustCom)},
  pages={952--959},
  year={2024},
  organization={IEEE}
}

@article{wei2024unified,
  title={Unified adversarial patch for visible-infrared cross-modal attacks in the physical world},
  author={Wei, Xingxing and Huang, Yao and Sun, Yitong and Yu, Jie},
  journal={IEEE Transactions on Pattern Analysis and Machine Intelligence},
  volume={46},
  number={4},
  pages={2348--2363},
  year={2024},
  publisher={IEEE}
}

@inproceedings{wei2023unified,
  title={Unified adversarial patch for cross-modal attacks in the physical world},
  author={Wei, Xingxing and Huang, Yao and Sun, Yitong and Yu, Jie},
  booktitle={Proceedings of the IEEE/CVF International Conference on Computer Vision},
  pages={4445--4454},
  year={2023}
}

@article{hu2025touap,
  title={Two-stage optimized unified adversarial patch for attacking visible-infrared cross-modal detectors in the physical world},
  author={Hu, Chengyin and Shi, Weiwen and Yao, Wen and Jiang, Tingsong and Tian, Ling and Li, Wen},
  journal={Applied Soft Computing},
  volume={168},
  pages={112678},
  year={2025},
  publisher={Elsevier}
}

@inproceedings{kim2022map,
  title={MAP: Multispectral adversarial patch to attack person detection},
  author={Kim, Taeho and Lee, Hyo Jin and Ro, Yong Man},
  booktitle={ICASSP 2022-2022 IEEE International Conference on Acoustics, Speech and Signal Processing (ICASSP)},
  pages={4853--4857},
  year={2022},
  organization={IEEE}
}

@article{sun2022dronevehicle,
  title={Drone-based RGB-Infrared cross-modality vehicle detection via uncertainty-aware learning},
  author={Sun, Yiming and Cao, Bing and Zhu, Pengfei and Hu, Qinghua},
  journal={IEEE Transactions on Circuits and Systems for Video Technology},
  volume={32},
  number={10},
  pages={6700--6713},
  year={2022},
  publisher={IEEE},
  doi={10.1109/TCSVT.2022.3168279}
}

@article{jia2021llvip,
  title={LLVIP: A visible-infrared paired dataset for low-light vision},
  author={Jia, Xinyu and Zhu, Chuang and Li, Minzhen and Tang, Wenqi and Zhou, Wenli},
  journal={arXiv preprint arXiv:2108.10831},
  year={2021},
  note={ICCV Workshop}
}

@article{zhu2020visdrone,
  title={Detection and tracking meet drones challenge},
  author={Zhu, Pengfei and Wen, Longyin and Du, Dawei and Bian, Xiao and Fan, Heng and Hu, Qinghua and Ling, Haibin},
  journal={IEEE Transactions on Pattern Analysis and Machine Intelligence},
  volume={44},
  number={11},
  pages={7380--7399},
  year={2020},
  publisher={IEEE},
  doi={10.1109/TPAMI.2021.3119563}
}

@article{redmon2018yolov3,
  title={YOLOv3: An incremental improvement},
  author={Redmon, Joseph and Farhadi, Ali},
  journal={arXiv preprint arXiv:1804.02767},
  year={2018}
}

@article{jocher2020yolov5,
  title={YOLOv5 by Ultralytics},
  author={Jocher, Glenn and others},
  journal={GitHub repository},
  year={2020},
  url={https://github.com/ultralytics/yolov5}
}

@article{jocher2023yolov8,
  title={YOLOv8 by Ultralytics},
  author={Jocher, Glenn and Chaurasia, Ayush and Qiu, Jing},
  journal={GitHub repository},
  year={2023},
  url={https://github.com/ultralytics/ultralytics}
}

@article{ren2015faster,
  title={Faster R-CNN: Towards real-time object detection with region proposal networks},
  author={Ren, Shaoqing and He, Kaiming and Girshick, Ross and Sun, Jian},
  journal={IEEE Transactions on Pattern Analysis and Machine Intelligence},
  volume={39},
  number={6},
  pages={1137--1149},
  year={2015},
  publisher={IEEE},
  doi={10.1109/TPAMI.2016.2577031}
}

@inproceedings{hu2021naturalistic,
  title={Naturalistic physical adversarial patch for object detectors},
  author={Hu, Yu-Chih-Tuan and Kung, Bo-Han and Tan, Daniel Stanley and Chen, Jun-Cheng and Hua, Kai-Lung and Cheng, Wen-Huang},
  booktitle={Proceedings of the IEEE/CVF international conference on computer vision},
  pages={7848--7857},
  year={2021}
}

@inproceedings{suryanto2023active,
  title={Active: Towards highly transferable 3d physical camouflage for universal and robust vehicle evasion},
  author={Suryanto, Naufal and Kim, Yongsu and Larasati, Harashta Tatimma and Kang, Hyoeun and Le, Thi-Thu-Huong and Hong, Yoonyoung and Yang, Hunmin and Oh, Se-Yoon and Kim, Howon},
  booktitle={Proceedings of the IEEE/CVF international conference on computer vision},
  pages={4305--4314},
  year={2023}
}

@inproceedings{abdelfattahadversarial,
  title={Adversarial attacks on camera-lidar models for 3d car detection. In 2021 IEEE},
  author={Abdelfattah, Mazen and Yuan, Kaiwen and Wang, Z Jane and Ward, Rabab},
  booktitle={RSJ International Conference on Intelligent Robots and Systems (IROS)},
  pages={2189--2194}
}

@inproceedings{wang2022fca,
  title={Fca: Learning a 3d full-coverage vehicle camouflage for multi-view physical adversarial attack},
  author={Wang, Donghua and Jiang, Tingsong and Sun, Jialiang and Zhou, Weien and Gong, Zhiqiang and Zhang, Xiaoya and Yao, Wen and Chen, Xiaoqian},
  booktitle={Proceedings of the AAAI conference on artificial intelligence},
  volume={36},
  number={2},
  pages={2414--2422},
  year={2022}
}

@inproceedings{suryanto2022dta,
  title={Dta: Physical camouflage attacks using differentiable transformation network},
  author={Suryanto, Naufal and Kim, Yongsu and Kang, Hyoeun and Larasati, Harashta Tatimma and Yun, Youngyeo and Le, Thi-Thu-Huong and Yang, Hunmin and Oh, Se-Yoon and Kim, Howon},
  booktitle={Proceedings of the IEEE/CVF Conference on Computer Vision and Pattern Recognition},
  pages={15305--15314},
  year={2022}
}

@article{duan2021learning,
  title={Learning coated adversarial camouflages for object detectors},
  author={Duan, Yexin and Chen, Jialin and Zhou, Xingyu and Zou, Junhua and He, Zhengyun and Zhang, Jin and Zhang, Wu and Pan, Zhisong},
  journal={arXiv preprint arXiv:2109.00124},
  year={2021}
}

@inproceedings{wang2021dual,
  title={Dual attention suppression attack: Generate adversarial camouflage in physical world},
  author={Wang, Jiakai and Liu, Aishan and Yin, Zixin and Liu, Shunchang and Tang, Shiyu and Liu, Xianglong},
  booktitle={Proceedings of the IEEE/CVF conference on computer vision and pattern recognition},
  pages={8565--8574},
  year={2021}
}

@article{hu2024physically,
  title={Physically structured adversarial patch inspired by natural leaves multiply angles deceives infrared detectors},
  author={Hu, Zhiyang and Yang, Xing and Zhao, Jiwen and Gao, Haoqi and Xu, Haoli and Mu, Hua and Wang, Yangyang},
  journal={Journal of King Saud University-Computer and Information Sciences},
  volume={36},
  number={6},
  pages={102122},
  year={2024},
  publisher={Elsevier}
}

@article{zhang2024pattern,
  title={Pattern corruption-assisted physical attacks against object detection in uav remote sensing},
  author={Zhang, Yu and Gong, Zhiqiang and Wen, Hao and Hu, Xikun and Xia, Xiaoyan and Jiang, Hejun and Zhong, Ping},
  journal={IEEE Journal of Selected Topics in Applied Earth Observations and Remote Sensing},
  volume={17},
  pages={12931--12944},
  year={2024},
  publisher={IEEE}
}

@inproceedings{xiang2025acattack,
  title={Acattack: Adaptive cross attacking rgb-t tracker via multi-modal response decoupling},
  author={Xiang, Xinyu and Yan, Qinglong and Zhang, Hao and Ma, Jiayi},
  booktitle={Proceedings of the Computer Vision and Pattern Recognition Conference},
  pages={22099--22108},
  year={2025}
}

@inproceedings{long2025cdupatch,
  title={Cdupatch: Color-driven universal adversarial patch attack for dual-modal visible-infrared detectors},
  author={Long, Jiahuan and Yao, Wen and Jiang, Tingsong and Hou, Jiacheng and Jia, Shuai and Wu, Junqi and Zhang, Xiaoya and Zheng, Xiaohu and Ma, Chao},
  booktitle={Proceedings of the 33rd ACM International Conference on Multimedia},
  pages={1462--1470},
  year={2025}
}

@article{han2025evolutionary,
  title={Evolutionary multiobjective cross-spectral adversarial attacks with synergistic patches},
  author={Han, Wencheng and Li, Hao and Gong, Maoguo and Wu, Yue and Qin, AK and Xing, Lining and Zhou, Yu},
  journal={IEEE Transactions on Systems, Man, and Cybernetics: Systems},
  year={2025},
  publisher={IEEE}
}

@inproceedings{wei2023physically,
  title={Physically adversarial infrared patches with learnable shapes and locations},
  author={Wei, Xingxing and Yu, Jie and Huang, Yao},
  booktitle={Proceedings of the IEEE/CVF conference on computer vision and pattern recognition},
  pages={12334--12342},
  year={2023}
}

@software{yolo11_ultralytics,
  author = {Glenn Jocher and Jing Qiu},
  title = {Ultralytics YOLO11},
  version = {11.0.0},
  year = {2024},
  url = {https://github.com/ultralytics/ultralytics},
  orcid = {0000-0001-5950-6979, 0000-0003-3783-7069},
  license = {AGPL-3.0}
}

@book{vollmer2018infrared,
  title={Infrared Thermal Imaging: Fundamentals, Research and Applications},
  author={Vollmer, Michael and M{\"o}llmann, Klaus-Peter},
  year={2018},
  publisher={John Wiley \& Sons},
  address={Hoboken, NJ},
  edition={2nd}
}

\appendix
\section*{Supplementary Material}

\section{Overview}
To enhance reproducibility and provide deeper insights into our proposed framework, this supplementary material includes:
\begin{itemize}
    \item \textbf{Algorithm of MEC-Patch.} We present a complete algorithmic description of MEC-Patch together with a plain-language explanation of the cross-modal execution logic.
    \item \textbf{In-depth Analysis of Adversarial Patches.} We examine patch configurations, show visible-infrared correspondence visualizations, and explain the rationale behind the selected shapes for physical material splicing.
    \item \textbf{Material Property Reference.} We provide a technical reference table listing visible-light color values and emissivity coefficients ($\epsilon$) for every industrial material used in our experiments.
    \item \textbf{Extended Attack Visualizations.} We include an expanded gallery of attack results and offer principle-based explanations of the emissivity-driven deception mechanism.
    \item \textbf{Discussion.} We conclude with a synthesized discussion on the overall effectiveness of the patches, their social impact, and promising directions for future research.
\end{itemize}
Each section reports empirical observations and provides analytical discussions that highlight the methodological advantages of our physics-grounded approach.

\section{Algorithm}
\label{sec:alg}
We formulate the optimization of MEC-Patch as a closed-loop evolutionary pipeline. Starting from a randomly initialized population of genotypes, each encoding the patch’s position, dartboard-discretized shape, and material assignments, we iteratively evaluate, resample, and evolve candidates over $G$ generations. In each generation the rendering engine synthesizes dual-modal adversarial images by mapping material IDs to RGB reflectivity for the visible branch and to emissivity-driven thermal radiation
\[
I = \mathcal{M}(\epsilon_i\cdot\sigma T^{4})
\]
for the infrared branch, then computes a multi-objective fitness vector $\Phi(\mathbf{g})$ that balances attack success, stealthiness, and area. To enhance cross-scene robustness, the Dynamic Adversarial Resampling (DAR) strategy adaptively reweights the scene pool according to per-scene attack difficulty, forcing the optimizer to focus on challenging environments. The NSGA-II engine subsequently drives the population toward the Pareto front via non-dominated sorting, selection, crossover, and mutation. Upon convergence we decode the genotype that attains the lowest attack loss into the final physical patch $P^{*}$. This pipeline guarantees physically consistent cross-modal perturbations while generalizing across diverse environmental conditions.

\begin{algorithm}[t]
\caption{MEC-Patch: Evolutionary Optimization}
\label{alg:mec_patch}
\SetAlgoLined
\KwIn{Detector $f(\cdot)$, scene set $\mathcal{S}$, initial weights $\mathbf{W}=[\omega_{1},\ldots,\omega_{S}]$,\\
\quad population size $N$, generations $G$, evaluation scenes $K$,\\
\quad emissivity library $\{\epsilon_{i}\}$ and RGB reflectance library}
\KwOut{Optimal patch $P^{*}$ with genotype $\mathbf{g}^{*}$}
\BlankLine
\textbf{Initialization:}\\
\For{$i\leftarrow 1$ \KwTo $N$}{
    Sample $\mathbf{g}_{\mathrm{pos}}$, $\mathbf{g}_{\mathrm{sha}}$ via $r(\theta)=\sqrt{(a\cos\theta)^{2}+(b\sin\theta)^{2}}$\;
    Discretize into dartboard sectors, assign $\mathbf{g}_{\mathrm{mat}}$\;
    Form $\mathbf{g}_{i}=[\mathbf{g}_{\mathrm{pos}},\mathbf{g}_{\mathrm{sha}},\mathbf{g}_{\mathrm{mat}}]$\;
}
$\mathcal{P}_{0}\leftarrow\{\mathbf{g}_{1},\ldots,\mathbf{g}_{N}\}$\;
\BlankLine
\For{$t\leftarrow 1$ \KwTo $G$}{
    \textbf{DAR Sampling:} Select $K$ scenes from $\mathcal{S}$ according to $\mathbf{W}^{t}$\;
    \BlankLine
    \textbf{Rendering \& Evaluation:}\\
    \For{each $\mathbf{g}\in\mathcal{P}_{t-1}$}{
        Render $(\mathcal{R}_{\mathrm{vis}},\mathcal{R}_{\mathrm{ir}})$ via RGB mapping and $I=\mathcal{M}(\epsilon_{i}\sigma T^{4})$\;
        Compute fitness $\Phi(\mathbf{g})=[f_{\mathrm{atk}},f_{\mathrm{stealth}},f_{\mathrm{area}}]^{\top}$\;
    }
    \BlankLine
    \textbf{DAR Update:} Update $\omega_{s}^{t+1}$ using attack error $\varepsilon_{s}^{t}$\;
    \BlankLine
    \textbf{NSGA-II:} Non-dominated sorting, tournament selection, SBX crossover, mutation\;
    Generate offspring $\mathcal{Q}_{t}$\;
    Merge $\mathcal{Q}_{t}$ with $\mathcal{P}_{t-1}$ and select the top $N$ individuals by crowding distance\;
    $\mathcal{P}_{t}\leftarrow$ selected population\;
}
\BlankLine
$\mathbf{g}^{*}\leftarrow\arg\min_{\mathbf{g}\in\mathcal{P}_{G}}f_{\mathrm{atk}}(\mathbf{g})$\;
$P^{*}\leftarrow\mathcal{R}(\mathbf{g}^{*})$\;
\Return{$P^{*}$}
\end{algorithm}

\section{MEC-Patch Analysis}

\begin{figure}[t]
    \centering
    \includegraphics[width=0.8\linewidth]{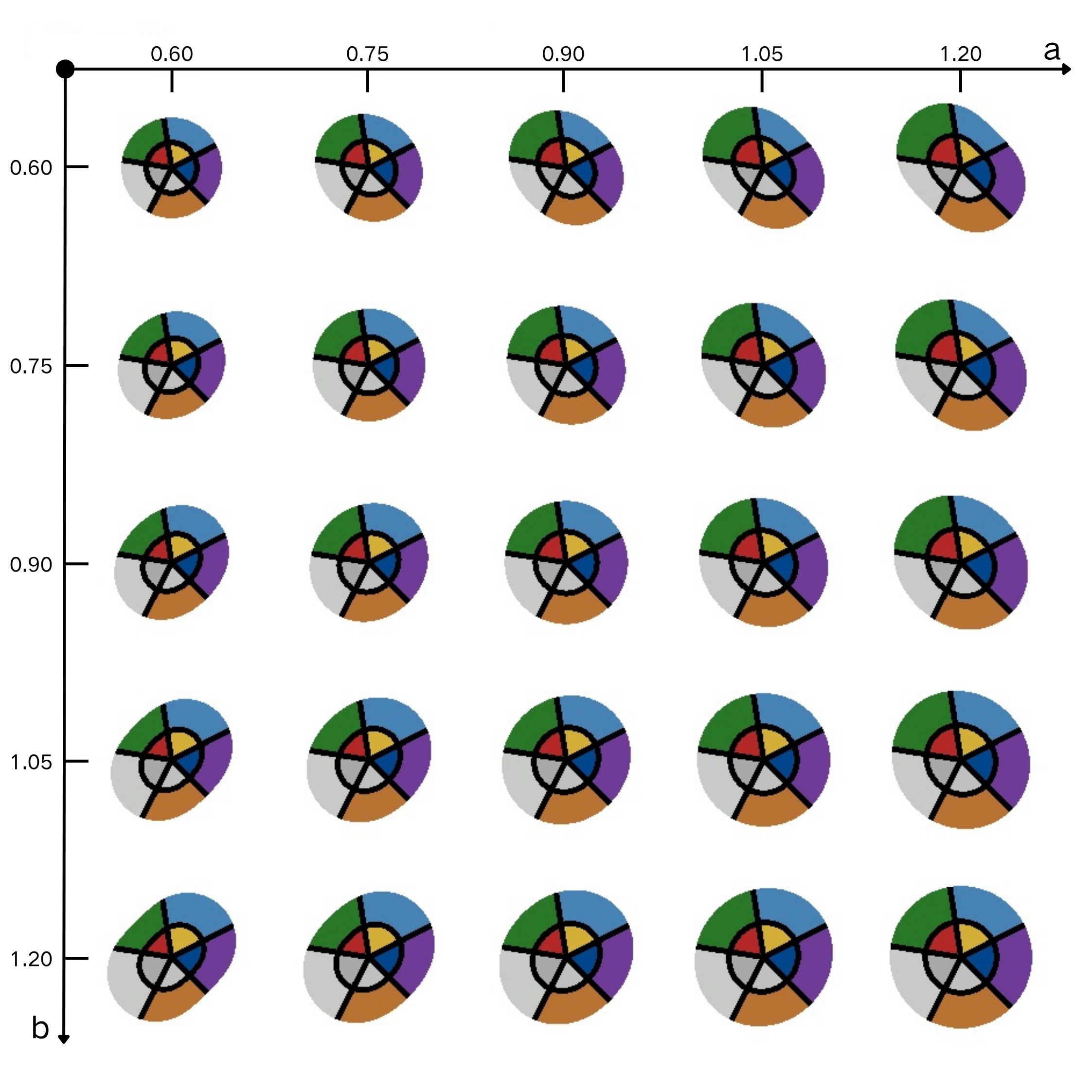}
    \caption{Parameterized shape space. Varying semi-axis parameters $(a,b)$ generates contours from circles to ellipses.}
    \label{fig:shape_space}
\end{figure}

\noindent\textbf{Parameterized shape space (Fig.~\ref{fig:shape_space}).} 
We vary the polar semi-axes $(a,b)$ to generate contours from circles to ellipses. This continuous parameterization allows NSGA-II to adapt the patch footprint to different vehicle surfaces.

\begin{figure}[b]
    \centering
    \includegraphics[width=0.8\linewidth]{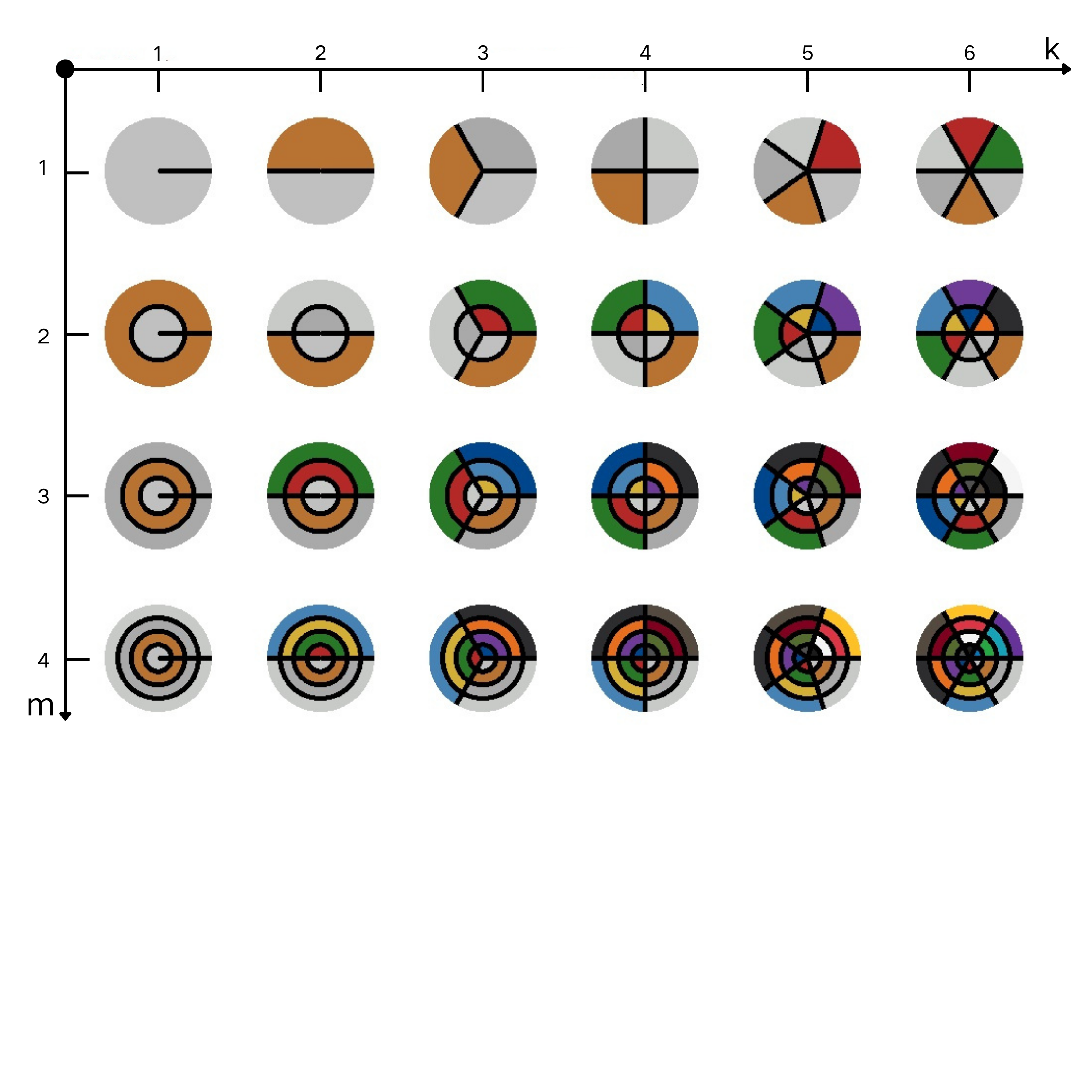}
    \vspace{-2cm}
    \caption{Dartboard topological discretization. The patch interior is partitioned into $K$ angular sectors and $m$ radial rings.}
    \label{fig:dartboard_topology}
\end{figure}

\noindent\textbf{Dartboard discretization (Fig.~\ref{fig:dartboard_topology}).} 
The interior is partitioned into $K$ angular sectors and $m$ radial rings, each assigned an independent material index. 
This structured grid transforms the discrete material selection problem into a form amenable to evolutionary optimization while keeping fabrication feasible (straight cuts and concentric arcs only).

\begin{figure}[t]
\centering
\includegraphics[width=\linewidth]{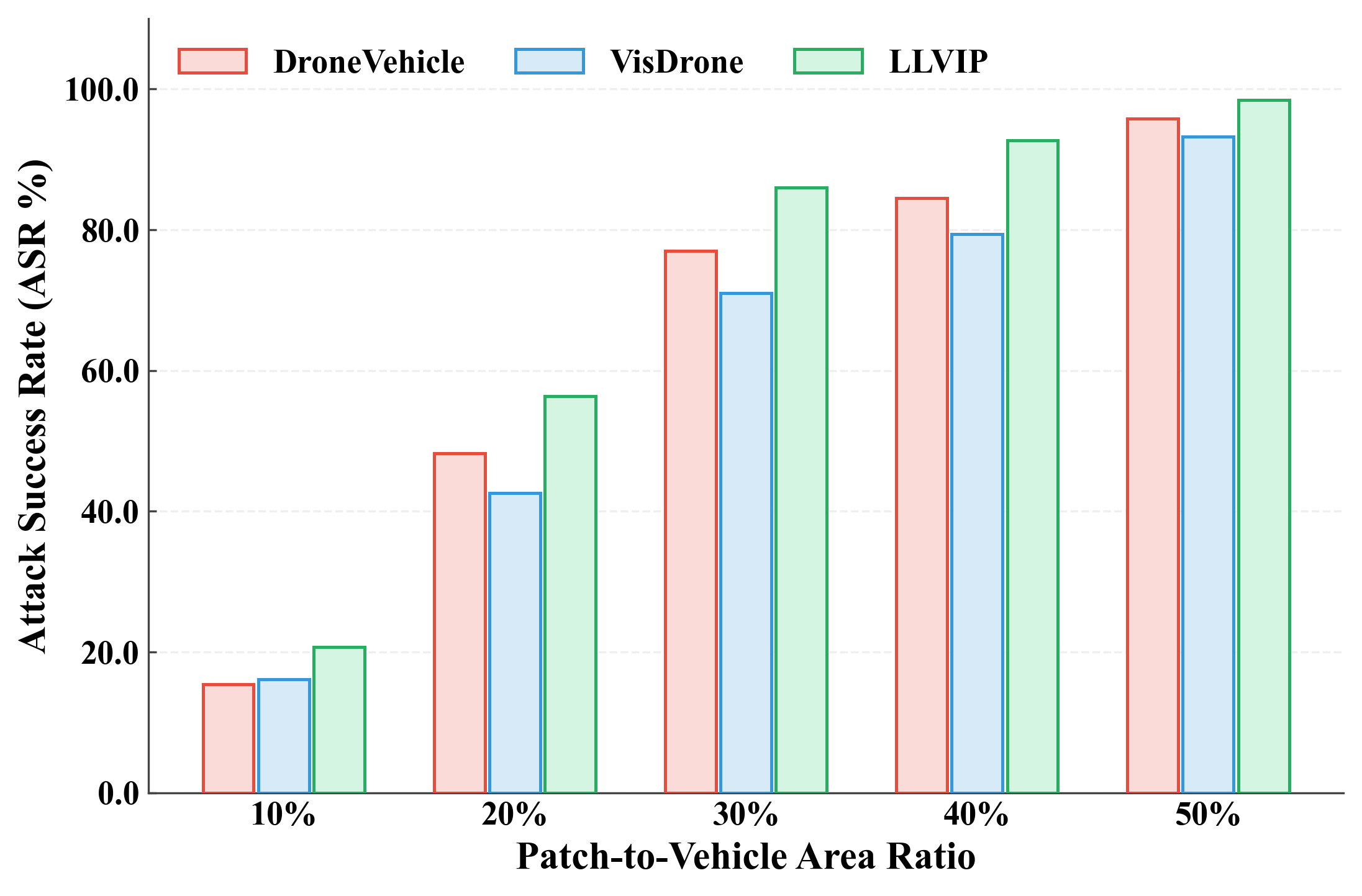}
\caption{ASR under different patch area ratios.}
\label{fig:area_ratio}
\end{figure}

\noindent\textbf{Sensitivity to Patch Area Ratio.}
Figure~\ref{fig:area_ratio} illustrates the ASR evolution across varying patch-to-vehicle area ratios (10\%--50\%). While a 10\% ratio provides insufficient adversarial energy, a sharp performance surge is observed at the 30\% threshold, where the ASR exceeds 70\% across all datasets. This identifies 30\% as the optimal equilibrium point, achieving high disruptive power while maintaining a strictly constrained footprint. As the ratio further increases to 50\%, the ASR approaches saturation (near 100\% on LLVIP), confirming that our MEC-based textures efficiently maximize feature disruption within limited physical areas.

\begin{figure}[h]
\centering
\includegraphics[width=0.85\linewidth]{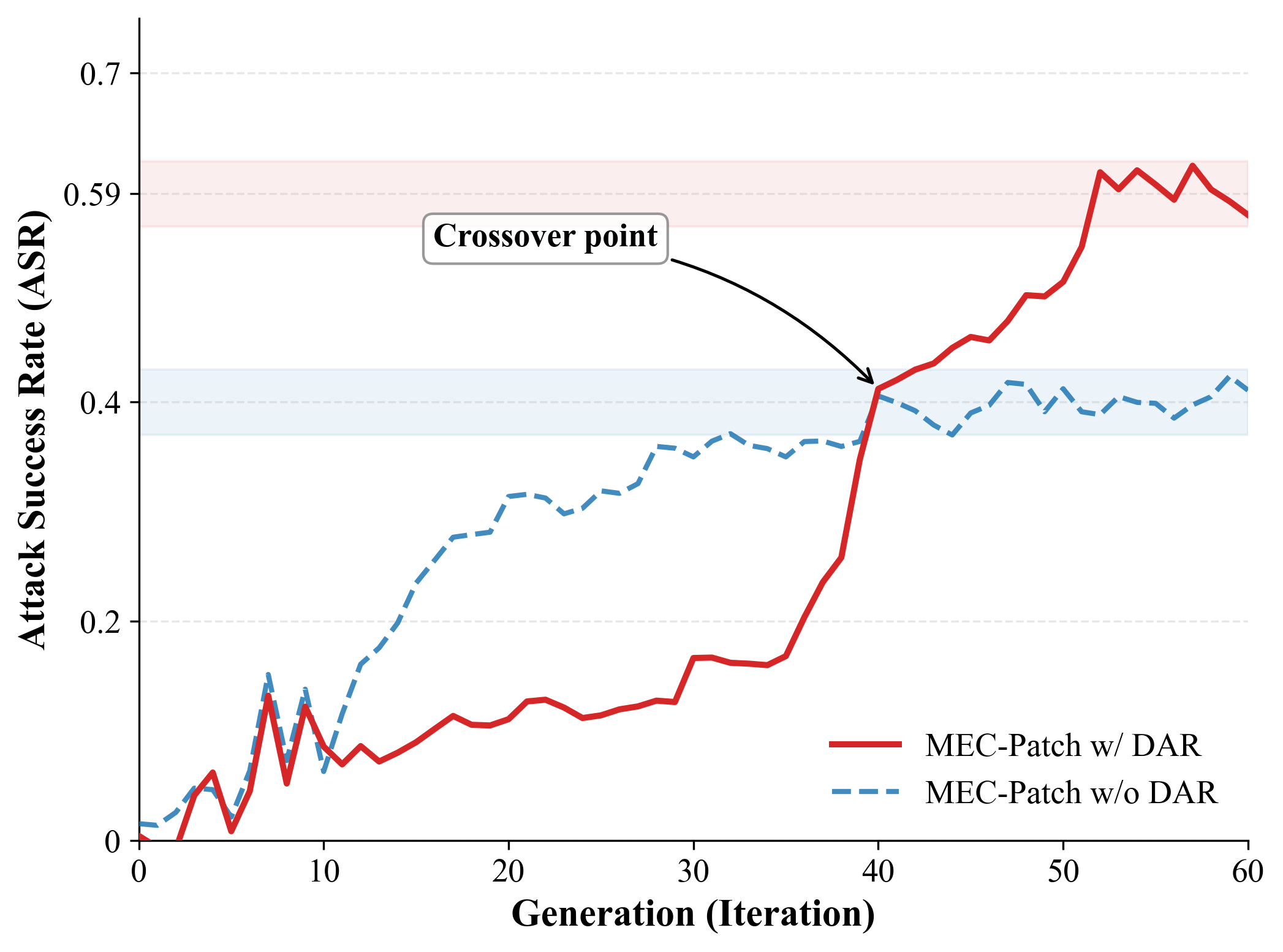}
\caption{Convergence curves comparing the DAR strategy with uniform sampling. DAR yields a late-stage breakthrough and higher final ASR.}
\label{fig:supp_dar_convergence}
\end{figure}

\noindent\textbf{Convergence Analysis of Dynamic Adversarial Resampling}
To further illustrate the effectiveness of the proposed Dynamic Adversarial Resampling (DAR) strategy, we compare its optimization trajectory against a uniform sampling baseline. As shown in Figure~\ref{fig:supp_dar_convergence}, uniform sampling quickly plateaus due to local optima. In contrast, the DAR-enhanced population, while initially exhibiting lower ASR by focusing on hard examples, achieves a distinct late-stage breakthrough and ultimately improves global ASR by approximately 15 percentage points. This confirms that adaptive scene weighting is critical for overcoming environmental imbalance and enhancing cross-scenario robustness.

\begin{figure}[t]
\centering
\includegraphics[width=\linewidth]{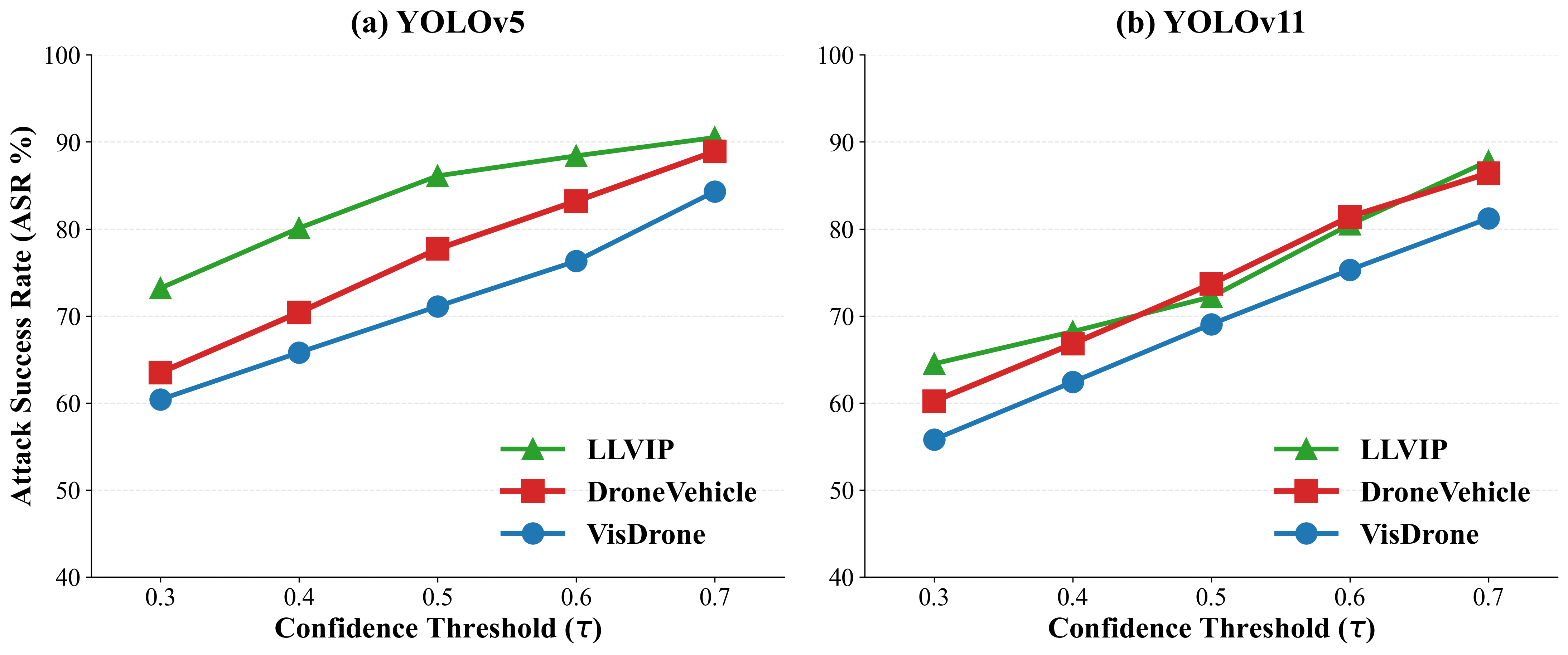}
\caption{ASR across different confidence thresholds.}
\label{fig:confidence_threshold}
\end{figure}

\noindent\textbf{Analysis of ASR across Confidence Thresholds.}
We evaluated the impact of varying detection thresholds ($\tau$). Figure~\ref{fig:confidence_threshold} shows that the proposed method maintains a high ASR even at a strict threshold of $\tau=0.3$. Analysis reveals that the patch causes a catastrophic collapse of the confidence space, typically suppressing the target's response to below 0.15--0.20. This confirms that our material-based attack induces fundamental feature misjudgments rather than minor score fluctuations.

\begin{figure}[htbp]
    \centering
    \includegraphics[width=0.8\linewidth]{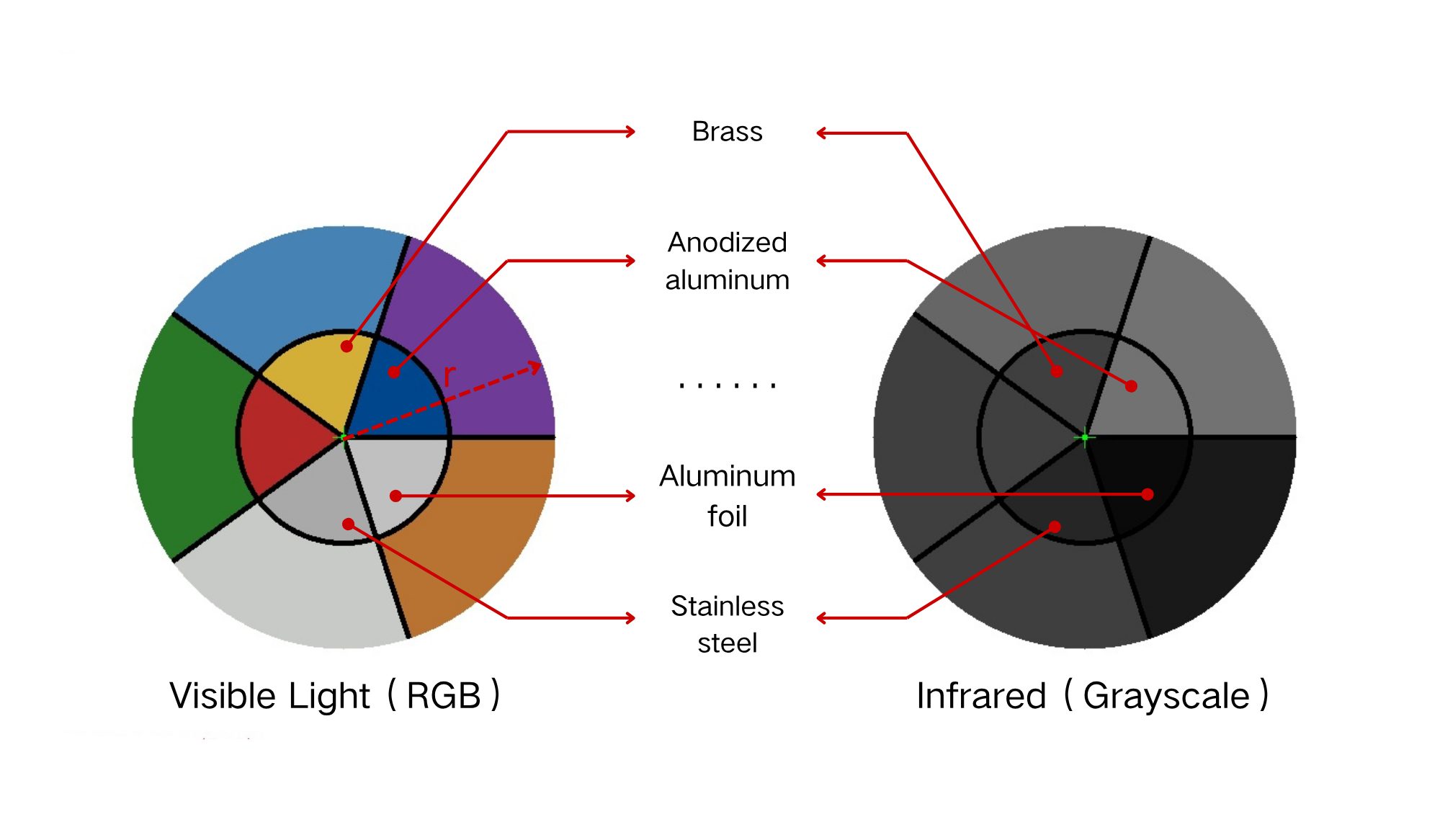}
    \vspace{-5mm}
    \caption{Material-level cross-modal response. Visible appearance (left) versus infrared emission (right) of each material cell, illustrating the cross-modal manifestation of the same physical material.}
    \label{fig:material_labeling}
\end{figure}

\noindent\textbf{Material-level cross-modal response (Fig.~\ref{fig:material_labeling}).} 
Exemplary cells are shown with their material types, visible appearance (left), and corresponding infrared emission (right), illustrating how each material manifests across the two modalities.
Materials are deliberately paired to decouple RGB reflectivity from thermal emissivity (e.g., dark-coated aluminum stays cold in LWIR while similarly coated ceramic stays hot), creating cross-spectral inconsistencies that disrupt visible-infrared cross-modal feature alignment.

\noindent\textbf{Dartboard design.} 
Three factors jointly explain its effectiveness. 
First, the discrete nature of physical material splicing prohibits gradient-based optimization; the dartboard grid provides a structured search space that NSGA-II can efficiently explore via crossover and mutation. 
Second, monolithic patches produce only uniform infrared shifts, which detectors can partially compensate for, whereas alternating low-$\epsilon$ and high-$\epsilon$ cells generate local emissivity discontinuities that accumulate adversarial gradient conflict.
Third, under the Stefan-Boltzmann law, ambient temperature changes globally scale $I \propto \epsilon \cdot T^4$ while preserving relative contrast, so the high-frequency pattern remains thermally stable across day-night cycles—explaining why MECPatch generalizes without retraining.

\section{Material Property Reference}
We provide a technical reference table (Table~\ref{tab:materials}) listing visible-light color values and emissivity coefficients $\epsilon$ for all 26 industrial materials used in our experiments.  
The library is designed to decouple RGB reflectivity from thermal emissivity—materials with similar visible appearances (e.g., dark-coated aluminum and ceramic) can exhibit opposite infrared signatures ($\epsilon \approx 0.25$ vs. $0.85$). 
This contrast forms the foundation of our dartboard-based adversarial patches, enabling high-frequency thermal patterns that remain stable under temperature variations.

We also note that physical implementation inevitably introduces minor discrepancies from ideal parameter values, such as variations in coating thickness or cutting precision. 
These imperfections do not compromise our physics-driven framework, as the attack's effectiveness depends primarily on the relative emissivity contrast between materials rather than absolute parameter accuracy.

\begin{table*}[t]
\centering
\caption{Visible-light color values and emissivity coefficients of industrial materials.}
\label{tab:materials}
\begin{tabular}{lccc}
\hline
\textbf{Material} & \textbf{RGB Color} & \textbf{Emissivity $\epsilon$} & \textbf{IR Signature} \\
\hline
\multicolumn{4}{c}{\textit{Low-emissivity metals ($\epsilon < 0.30$)}} \\
\hline
Aluminum foil & (192,192,192) & 0.04 & Very cold \\
Brass (unoxidized, polished) & (184,115,51) & 0.06 & Very cold \\
Stainless steel & (169,169,169) & 0.15 & Cold \\
Anodized Al (5$\mu$m, natural) & (200,202,200) & 0.25 & Cold \\
Anodized Al (5$\mu$m, red/green/gold) & (180,40,40) / (40,120,40) / (212,175,55) & 0.25 & Cold \\
\hline
\multicolumn{4}{c}{\textit{Medium-emissivity anodized Al ($0.40 \leq \epsilon < 0.70$)}} \\
\hline
Anodized Al (10$\mu$m, blue) & (70,130,180) & 0.40 & Medium-cold \\
Anodized Al (12$\mu$m, dark blue/purple/orange) & (0,70,140) / (110,60,150) / (230,110,30) & 0.45 & Medium \\
Anodized Al (15$\mu$m, black) & (45,45,48) & 0.55 & Medium-warm \\
Anodized Al (20$\mu$m, dark gray/olive/burgundy) & (80,80,80) / (85,107,47) / (128,0,32) & 0.65 & Warm \\
\hline
\multicolumn{4}{c}{\textit{High-emissivity materials ($\epsilon \geq 0.70$)}} \\
\hline
Hard anodized Al (30$\mu$m, brown) & (84,74,63) & 0.70 & Warm \\
Hard anodized Al (50$\mu$m, deep black) & (28,28,28) & 0.82 & Hot \\
Ceramic (white/red/yellow/blue/green/cyan/purple) & Various white/bright colors & 0.84--0.86 & Hot \\
Graphite & (33,33,33) & 0.70--0.95 & Very hot \\
Rubber & (52,58,64) & 0.90 & Very hot \\
\hline
\end{tabular}
\end{table*}

\section{Extended Attack Visualizations}
\begin{figure*}[t]
    \centering
    \includegraphics[width=1.0\linewidth]{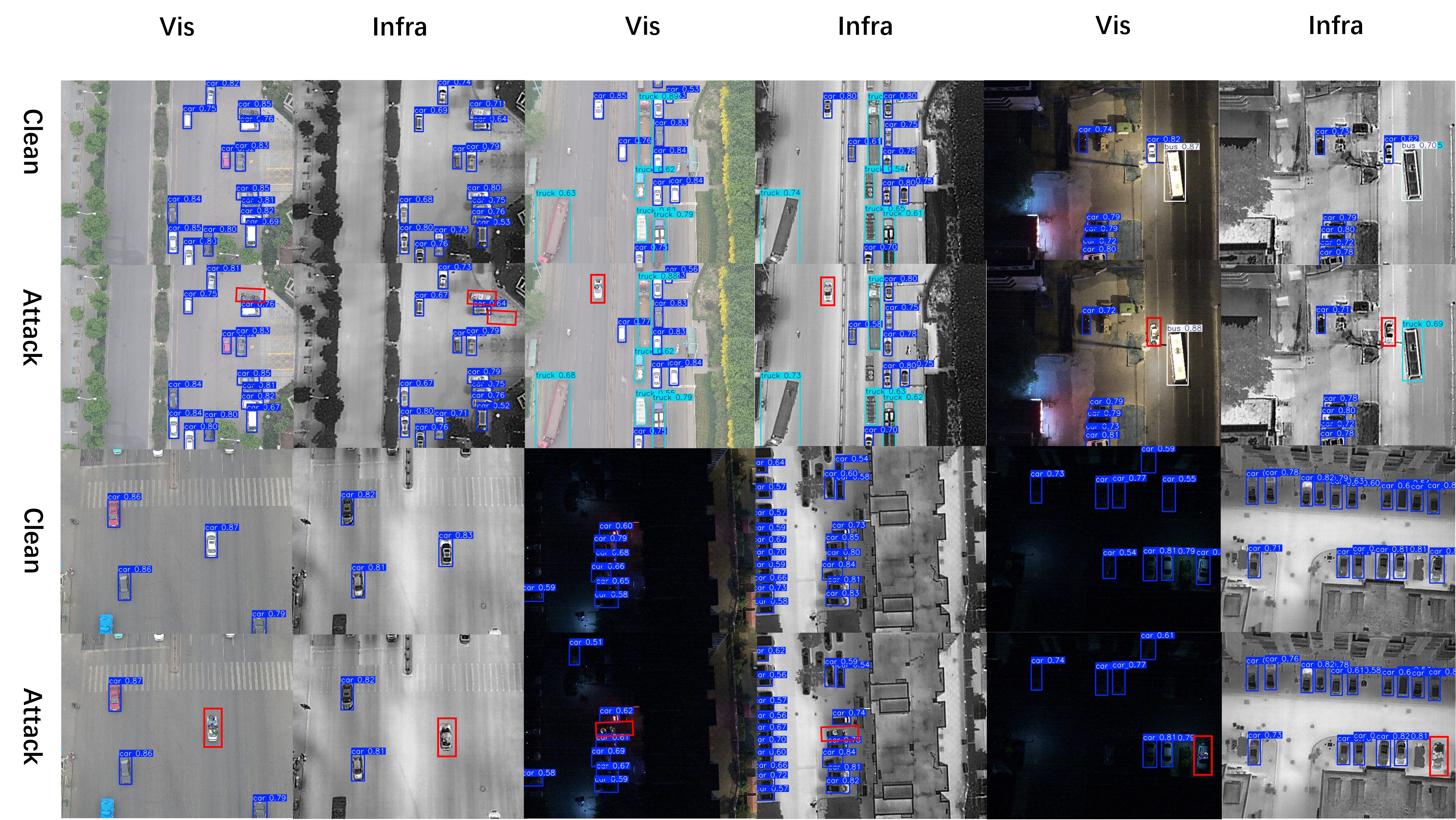}
    \caption{Extended attack visualizations of MEC-Patch across diverse scenes.}
    \label{fig:more_vis}
\end{figure*}

We provide additional visualization results (Fig.~\ref{fig:more_vis}) demonstrating that MEC-Patch consistently produces high-fidelity visible appearances and contrasting infrared signatures across diverse scenes, confirming the effectiveness of the emissivity-driven deception mechanism.

\section{Discussion}
We have demonstrated that MEC-Patch provides a physics-grounded approach to visible-infrared cross-modal attacks, with key advantages including thermal consistency via the Stefan-Boltzmann formulation, discrete optimization of physically realizable patches via dartboard-based material encoding, and environmental robustness via the DAR strategy. 
We also acknowledge that several limitations remain: the fabrication precision required for high-frequency dartboard patterns is susceptible to real-world manufacturing tolerances, which may cause minor deviations from the ideal design, and the current material library of 26 industrial materials, while sufficiently diverse for proof-of-concept, is not exhaustive.

This study is conducted strictly for academic research purposes to identify potential security flaws in multimodal perception systems and will not be applied to any commercial use. 
By uncovering these vulnerabilities, our work provides a diagnostic benchmark to catalyze the realization of security-by-design architectures in autonomous systems.

For future work, we plan to extend our material library to include additional industrial coatings, fabrics, and paints to further enrich the adversarial design space. 
We also aim to transition from 2D planar approximations to high-fidelity 3D adversarial modeling that accounts for complex geometric curvatures, integrating parametric mesh deformation with physical rendering engines to simulate more realistic physical deployment scenarios, thereby establishing more rigorous security assessment protocols for the next generation of reliable multimodal perception systems.

\end{document}